\documentclass[%
 reprint,
 amsmath,amssymb,
 aps,prc,
 floatfix
]{revtex4-2}

\usepackage{graphicx}
\usepackage{bm}
\usepackage{braket}
\usepackage{siunitx}
\usepackage[unicode]{hyperref}

\newcommand{\aton}{\alpha\text{-}{}^2n}
\newcommand{\atoa}{\alpha \text{-} \alpha}
\newcommand{\nton}{{}^2n \text{-} {}^2n}
\newcommand{\He}{^8\mathrm{He}}
\newcommand{\C}{^{12}\mathrm{C}}
\newcommand{\Be}{^{10}\mathrm{Be}}

\begin{document}
 \title{3-body cluster gas structures in the excited $0^+$ states of $N=6$ nuclei}
 \author{Kosei Nakagawa}
\affiliation{Department of Physics, Kyoto University, Kyoto, 606-8502, Japan}
\author{Yoshiko Kanada-En'yo}
\affiliation{Department of Physics, Kyoto University, Kyoto, 606-8502, Japan}
\affiliation{Yukawa Institute for Theoretical Physics, Kyoto University, Kyoto, 606-8502, Japan}

\date{\today}

\begin{abstract}
  \begin{description}
    \item[Background] Dineutron clustering in $\He(0_2^+)$ has been attracting interest and
its analogy with  the $3\alpha$ clustering in $\C(0_2^+)$ has been discussed in theoretical and experimental
studies.
    \item[Purpose]
We investigate the  cluster structure of $\alpha+{}^2n+{}^2n$ in $\He(0_2^+)$ and compare it with those of $2\alpha+{}^2n$ in $\Be(0_2^+)$ and $3\alpha$ in  $\C(0_2^+)$ to understand the emergence mechanism of
the 3-body cluster gas states.
    \item[Methods] We apply an extended cluster model combined with cluster breaking components
in a microscopic framework using effective nuclear forces based on nucleon degrees of freedom
including the antisymmetrization between nucleons.
    \item[Results]
In detailed analysis of the 3-body cluster structures of excited $0^+$ states, it is shown that
$\He(0_2^+)$ exhibits 3-body cluster gas feature of $\alpha+{}^2n+{}^2n$ similar to $\C$ but
contains significant mixing of a 2-body-like $(\alpha+{}^2n)+{}^2n$ component.
We discuss cluster structures of $\He(0_2^+)$ and $\C(0_2^+)$ from the point of view of
inter-cluster energies of $\nton$ and $\aton$
in comparison with the $\atoa$ energies and find that the origin for
the 2-body-like $(\alpha+{}^2n)+{}^2n$ mixing is the unbalance of the $\aton$ and
$\nton$ energies, in which
Pauli effects play an essential role through the kinetic energy loss and internal potential energy loss of clusters.
\item[Conclusions]
We clarify the emergence mechanism of
the 3-body cluster gas state in excited $0^+$ states of $\He$ and $\C$ systems.
The balance of inter-cluster interactions is essential for the appearance of 3-body cluster gas states.
Microscopic effects, \textit{i.e.}, the Pauli effects of nucleons between clusters play a crucial role
in cluster structures of excited $0^+$ states.
  \end{description}
\end{abstract}

\maketitle

\section{introduction}
Light nuclei often have cluster states, especially in their excited states near threshold energy~\cite{IkedaRule}.
The $0_2^+$ state of $\C$, which is known as the Hoyle state, is a typical example of a developed cluster state of three $\alpha$ clusters.
Its excitation energy $\SI{7.65}{MeV}$~\cite{kelley2017energy} is located $\SI{0.380}{MeV}$ below the $3\alpha$ energy threshold, which nicely follows the energy threshold rule~\cite{IkedaRule}.
This state is understood as a cluster-gas of weakly interacting $3\alpha$ clusters in $S$-wave motions~\cite{kanada2007C,suhara2015,otsuka2022aclustering}, and characterized by a strong isoscalar monopole (IS0) transition from the ground state~\cite{yamada2008,PhysRevLett.105.022501}.
Tohsaki \textit{et al.} stressed bosonic behaviors of $3\alpha$ clusters and proposed a concept of $\alpha$ condensations~\cite{THSR}.
Existence of cluster-gas states in other nuclei such as ${}^{11}\rm{B}$, ${}^{11}\rm{C}$, and ${}^{13}\rm{C}$ has attracted attention in terms of Hoyle-analogue states~\cite{kanada2007negative,PhysRevC.91.014316,PhysRevC.101.024317}, and is investigated also for ${}^{16}\rm{O}$ and ${}^{20}\rm{Ne}$ by applying $\alpha$-condensation models~\cite{THSR,zhou2023condensate}.

Meanwhile, people discussed the existence of exotic clusters near energy threshold in neutron-rich nuclei.
Especially, pairs of neutrons exhibit spatial correlations at the surface of nuclei, and are called "dineutrons (${}^2n$)", which can be formed even in stable nuclei~\cite{catara1984,pillet2007}.
Dineutrons play important roles in structures of nuclei near the neutron dripline, including $\He$.
In the ground state of $\He$, four valence neutrons outside the $\alpha$ core form spin-zero pairs on the nuclear surface~\cite{hagino2005pairing}.
One of the authors (Y. K.) pointed out the dineutron formation and its clustering behavior in the excited state, $\He(0_2^+)$~\cite{Kanada2007Structure}.
The dineutron formation induces a bosonic behavior of dineutron clusters, and causes a condensation of dineutrons in $\He(0_2^+)$, which is analogous to the $3\alpha$ condensation in $\C(0_2^+)$, as discussed in theoretical works~\cite{kobayashi2013}.
Yang \textit{et al.}~\cite{yang2024} discovered the $0_2^+$ state of $\He$ at the excitation energy $E_x(0_2^+)=\SI{6.66}{MeV}$, which is $\SI{3.54}{MeV}$ above the $\alpha+4n$ energy threshold, and verified contributions of the $\alpha+{}^2n+{}^2n$ clusters through the observation of strong IS0 transition between the $0_1^+$ and $0_2^+$ states, as in the case of $\C(0_2^+)$.

In our previous research~\cite{nakagawa2025alpha}, we constructed a model wave function for $\He$ that incorporated the $\alpha+{}^2n+{}^2n$ 3-cluster correlations and breaking of ${}^2n$ clusters, and successfully reproduced the observed $E_x(0_2^+)$ and IS0 transition strength of $\He$.
We discussed the $\alpha+{}^2n+{}^2n$ cluster structure in $\He(0_2^+)$ and found cluster-gas features, namely wide spatial distributions and $S$-wave motions of the three clusters, similar to those of the $3\alpha$ cluster-gas in $\C(0_2^+)$.
Although the 3-body cluster-gas component is dominant in $\He(0_2^+)$, we also found significant mixing with a 2-body-like $(\alpha+{}^2n)+{}^2n$ component.
This mixing indicates that the $\alpha+{}^2n$ correlation still remains in the $\alpha+{}^2n+{}^2n$ system, in contrast to the cluster-gas limit of the $3\alpha$ system in $\C(0_2^+)$.
We also investigated the $2\alpha+{}^2n$ cluster structure in ${}^{10}\mathrm{Be}(0_2^+)$ and found that it does not show clear 3-body cluster-gas nature.

In this work, we focus on the 3-cluster gas structure characterized by wide spatial distributions of three clusters in $S$-wave-like motions.
To understand universal features of 3-cluster gas states, we perform detailed analyses of the $\alpha+{}^2n+{}^2n$ cluster structure in $\He(0_2^+)$ within a microscopic framework based on nucleon degrees of freedom, as in our previous work~\cite{nakagawa2025alpha}, and compare its features with those of $3\alpha$ in $\C(0_2^+)$ and $2\alpha+{}^2n$ in ${}^{10}\mathrm{Be}(0_2^+)$.
We address two questions: what causes the differences in 3-body cluster structures among the three systems, $3\alpha$, $\alpha+{}^2n+{}^2n$, and $2\alpha+{}^2n$; and what conditions are essential for the emergence of cluster-gas states, particularly in non-identical cluster systems such as $\He(0_2^+)$.
To answer these questions, we focus on the inter-cluster interactions ($\alpha$-$\alpha$, $\alpha$-${}^2n$, and ${}^2n$-${}^2n$), which are crucial in determining the 3-body cluster structures of these systems.
These inter-cluster interactions originate from nucleon-nucleon interactions between composite clusters, where nucleon antisymmetrization, \textit{i.e.}, Pauli-blocking effects, plays an important role and can produce non-trivial effects specific to 3-cluster systems, unlike in 3-body systems of structureless particles.
We investigate the inter-cluster interactions from a microscopic viewpoint based on nucleon degrees of freedom to clarify the role of Pauli blocking.
We also perform test calculations by artificially changing the $\alpha$-${}^2n$ and ${}^2n$-${}^2n$ interactions to examine whether the 3-body cluster-gas component in the $0_2^+$ states can be enhanced in the $\alpha+{}^2n+{}^2n$ and $2\alpha+{}^2n$ systems.
For further analyses, we investigate 2-nucleon density distributions and discuss 2-cluster correlations in $\He$.

This paper is organized as follows.
In Sec.~\ref{sec:form}, our model wave functions and Hamiltonians are explained.
In Sec.~\ref{sec:res}, we analyze inter-cluster interactions to discuss the $\alpha+{}^2n+{}^2n$ cluster-gas state and discuss their role in 3-cluster systems. 
Sec.~\ref{sec:dis} is devoted to more detailed analyses of the 3-body cluster structures.
Finally, a summary and an outlook are given in Sec.~\ref{sec:sum}.

\section{formulation}
\label{sec:form}
In our previous study~\cite{nakagawa2025alpha} of $\alpha+{}^2n+{}^2n$ cluster structures in $\He(0_2^+)$, we applied the microscopic model in which wave functions and Hamiltonian are expressed from nucleon degrees of freedom.
The 3-body cluster correlations are taken into account through the cluster generator coordinate method (GCM) and the ${}^2n$ cluster breaking configurations are mixed.
The model wave function is expressed by a linear combination of Brink-Bloch (BB) cluster wave functions and shell-model wave functions~\cite{BBwf}.
In the present work, we adopt the same framework to perform detailed analyses of the $\alpha+{}^2n+{}^2n$ and $2\alpha+{}^2n$ cluster structures in $\He$ and $\Be$.
\subsection{model wave function}
The 3-body cluster wave function of $C_i \quad (i=1,2,3)$ with mass numbers $A_i$ centering at $\bm{R}_i$ is expressed as
\begin{equation}
  \label{3body}
  \begin{split}
    &\ket{C_1+C_2+C_3;\bm{R}_1,\bm{R}_2,\bm{R}_3}  \\
    &= {\cal A}  \left[ \ket{C_1;\bm{R}_1}  \ket{C_2;\bm{R}_2} \ket{C_3;\bm{R}_3} \right],
  \end{split}
\end{equation}
where $\ket{C_i;\bm{R}_i}$ is the $i$th cluster wave function and ${\cal A}$ is the antisymmetrizer.
In the 3-body configurations of the $\He$ wave function, $C_i$s are $\alpha$ and ${}^2n$ clusters, which are expressed by the harmonic oscillator (h.o.) $(0s)^4$ and $(0s)^2$ configurations centering at $\bm{R}_i$, respectively.

The 2-body cluster BB wave functions are defined as
\begin{equation} \label{2body}
  \ket{C_1+C_2;\bm{R}_1,\bm{R}_2}  = {\cal A}  \left[ \ket{C_1;\bm{R}_1} \ket{C_2;\bm{R}_2} \right].
\end{equation}
To express the one-dineutron breaking configurations in the 2-body cluster components of the $\He$ wave function, $C_1$ is chosen as the h.o. $(0s_{1/2})^4(0p_{3/2})^2$ cluster (simply denoted as ``$^6\rm{He}$'' in this paper) and $C_2$ is a ${}^2n$ cluster.
For analysis of inter-cluster interactions, $(C_1, C_2)$ are $(\alpha,\alpha)$, $(\alpha,{}^2n)$, $({}^2n,{}^2n)$ clusters with h.o. $(0s)^4$ or $(0s)^2$ configuration.

The 2-dineutron breaking configuration is taken into account to the neutron $p_{3/2}$-closure configuration in the $jj$-coupling shell model, which is denoted by $\ket{\Psi^{SM}}$.

The 3-body and 2-body configurations are projected onto the total angular momentum and parity $J^\pi=0^+$ eigenstates.
Furthermore, to exactly remove the center-of-mass motion of the total $A$-nucleon system of the $k$-body cluster wave functions ($k=2,3$), we set the center-of-mass of the cluster center parameters at the origin.

Practically, the $(0s_{1/2})^4(0p_{3/2})^2$, and $p_{3/2}$-closure configurations are expressed by means of the antisymmetrized quasicluster model~\cite{aqcm, aqcmganso, mono10Be9Li}.
In the present calculation, the width parameter $\nu=\SI{0.235}{fm^{-2}}$ is adopted as was done in the previous research~\cite{Kanada2007Structure,suhara2015}.
It corresponds to the h.o. energy quanta $\hbar \omega = \frac{2 \hbar^2 \nu}{m}=\SI{19.49}{MeV}$, where $\omega$ is the h.o. frequency, and $m$ is the nucleon mass.

For the 3-body cluster configurations, we use hyperspherical coordinates~\cite{descouvemont2019,suno2015,hsc2004} for the center positions $\bm{R}_i$ of three clusters $C_i$ ($i=1,2,3$) in Eq.~\eqref{3body}.
The scaled Jacobi coordinates $\bm{X}$ and $\bm{Y}$ are defined as
\begin{equation}
  \begin{split}
    & \bm{X} = \sqrt{\frac{A_1A_{2}}{A_{12}}} \left(\bm{R}_1-\bm{R}_2\right) , \quad \\
 & \bm{Y} = \sqrt{\frac{A_{12}A_3}{A}} \left(\frac{A_1\bm{R}_1+A_2\bm{R}_2}{A_{12}}-\bm{R}_3\right),
  \end{split}
\end{equation}
with $A_{12}=A_1+A_2$.
Then, the intrinsic 3-body spatial configurations are expressed by $3$ parameters,
$R \equiv \sqrt{\bm{X}^2+\bm{Y}^2}$, $\gamma \equiv \arctan(\frac{|\bm{X}|}{|\bm{Y}|})$, and angle $\theta$ (the angle between $\bm{X}$ and $\bm{Y}$), which are useful to express 3-body configurations, and now we can rewrite Eq.~\eqref{3body} as
\begin{equation}
  \ket{C_1+C_2+C_3;\bm{R}_1,\bm{R}_2,\bm{R}_3} \to \ket{C_1+C_2+C_3;R,\gamma,\theta}.
\end{equation}

We also rewrite the ${}^6\mathrm{He} + {}^2n$ 2-body wave functions in Eq.~\eqref{2body} with the inter-cluster distance as
\begin{equation} \label{2body_int}
  \ket{{}^6\mathrm{He}+{}^2n;\bm{R}_1,\bm{R}_2}  \to  \ket{{}^6\mathrm{He}+{}^2n;D_{{}^6\mathrm{He} \text{-} ^2n}}
\end{equation}
with $D_{{}^6\mathrm{He} \text{-} ^2n}=|\bm{R}_1-\bm{R}_2|$.

The total wave function of the $j$-th $0^+$ state of $\He$ is expressed by linear combinations of 3-cluster, 2-cluster, and SM wave functions as
\begin{equation} \label{wf8he}
  \begin{split}
    &\ket{\He(0_{j}^+)} \\
    =&\int dR \, d\gamma \, d\theta \,
    \hat{P}^{0+} c_{j}^{3B}(R,\gamma,\theta) \ket{\alpha+ {}^2n+ {}^2n;R,\gamma,\theta} \\
    +&\int dD_{{}^6\mathrm{He} \text{-} ^2n} c_{j}^{2B}(D_{{}^6\mathrm{He} \text{-} ^2n}) \hat{P}^{0+}\ket{^6\mathrm{He}+{}^2n;D_{{}^6\mathrm{He} \text{-} ^2n}}  \\
    +& c_{j}^{SM} \ket{\Psi^{SM}},
  \end{split}
\end{equation}
where $\hat{P}^{0+}$ is the projection operator onto the $0^+$ state, and the coefficients $c_{j}^{3B}(R,\gamma,\theta), \, c_{j}^{2B}(D_{{}^6\mathrm{He} \text{-} ^2n}), \, c_{j}^{SM}$ are determined by diagonalizing the norm and the Hamiltonian matrices.
The integrations are numerically calculated by summation of numerical mesh points of the generator coordinates.

Similarly, the $\Be$ wave function is expressed as
\begin{equation}
  \begin{split}
    &\ket{\Be(0_{j}^+)}\\
    =&\int dR \, d\gamma \, d\theta \,
    \hat{P}^{0+} c_{j}^{3B}(R,\gamma,\theta) \ket{\alpha+ {}^2n+ {}^2n;R,\gamma,\theta} \\
    &+\int da \, c_{j}^{2B}(a) \hat{P}^{0+}\ket{^6\mathrm{He}+\alpha;a},
  \end{split}
\end{equation}
where  $C_1=C_2=\alpha$, $C_3={}^2n$ is chosen in the $2\alpha+{}^2n$ configurations.

\subsection{Hamiltonian}
The Hamiltonian is composed of the kinetic energy $t_i$ of the $i$th nucleon, the effective nuclear forces including the central force $V_{ij}^{c}$ and the spin-orbit force $V_{ij}^{ls}$, and the Coulomb force $V_{ij}^{coul}$ between nucleons $i$ and $j$:
\begin{equation}
  \begin{split}
  H = T + U + U^{coul},  \\
  T = \sum_{i=1}^A t_i - t_G, \\
  U = \sum_{i<j} (V_{ij}^{c} + V_{ij}^{ls}) \\
  U^{coul} = \sum_{i<j} V_{ij}^{coul},
  \end{split}
\end{equation}
where $t_G$ is the kinetic energy of the total center of mass.
As for the central force, we use the Volkov No.2 force~\cite{VOLKOV196533}
\begin{equation} \label{Volkov}
  V_{ij}^{c} = (W-BP^{\sigma}-HP^{\tau}+MP^{\sigma}P^{\tau}) \sum_{k=1}^2 V_k e^{-(r_{ij} / b_k)^2},
\end{equation}
where $W$, $B$, $H$, and $M$ correspond to the amount of Wigner, Bartlett, Heisenberg, and Majorana terms, and $P^{\sigma}(P^{\tau})$ is the spin (isospin) exchange operator between particles $i$ and $j$.
Variable interaction parameters $W$, $B$, $H$, and $M$ are given as $M=0.58$, $W=0.42$, $B=H=0$ for $\He$ calculations, and $M=0.60$, $W=0.40$, $B=H=0$ for $\Be$, which follows our previous research~\cite{nakagawa2025alpha}.
The spin-isospin part of central forces can be decomposed into the channels of 2-nucleon spin-parity combinations: ${}^1E$ (singlet-even), ${}^3E$ (triplet-even), ${}^1O$ (singlet-odd), and ${}^3O$ (triplet-odd) as
\begin{equation}  \label{cent.equiv}
  \begin{split}
  V_{ij}^{c} =& \left\{ V_{1E}P({}^1E)+V_{3E}P({}^3E)+V_{1O}P({}^1O)+V_{3O}P({}^3O) \right\} \\
  & \qquad \times \sum_{k=1}^2 V_k e^{-(r_{ij} / b_k)^2}.
  \end{split}
\end{equation}
where $P({}^1E)$, $P({}^3E)$, $P({}^1O)$, and $P({}^3O)$ are projection operators onto each channel, and $V_{1E}$, $V_{3E}$, $V_{1O}$ and $V_{3O}$ are the strength factors of each component determined by the linear transformation of $W$, $M$, $H$, $B$.
For the spin-orbit interaction, we adopt the G3RS interaction~\cite{G3RSyamaguchi,G3RStamagaki} with the strength parameter $u_1=-u_2=\SI{1600}{MeV}$ as in our previous research~\cite{nakagawa2025alpha}.

\section{results} \label{sec:res}
\subsection{inter-cluster interaction} \label{sec.res.int}
\begin{figure}
  \includegraphics[width=5.5cm]{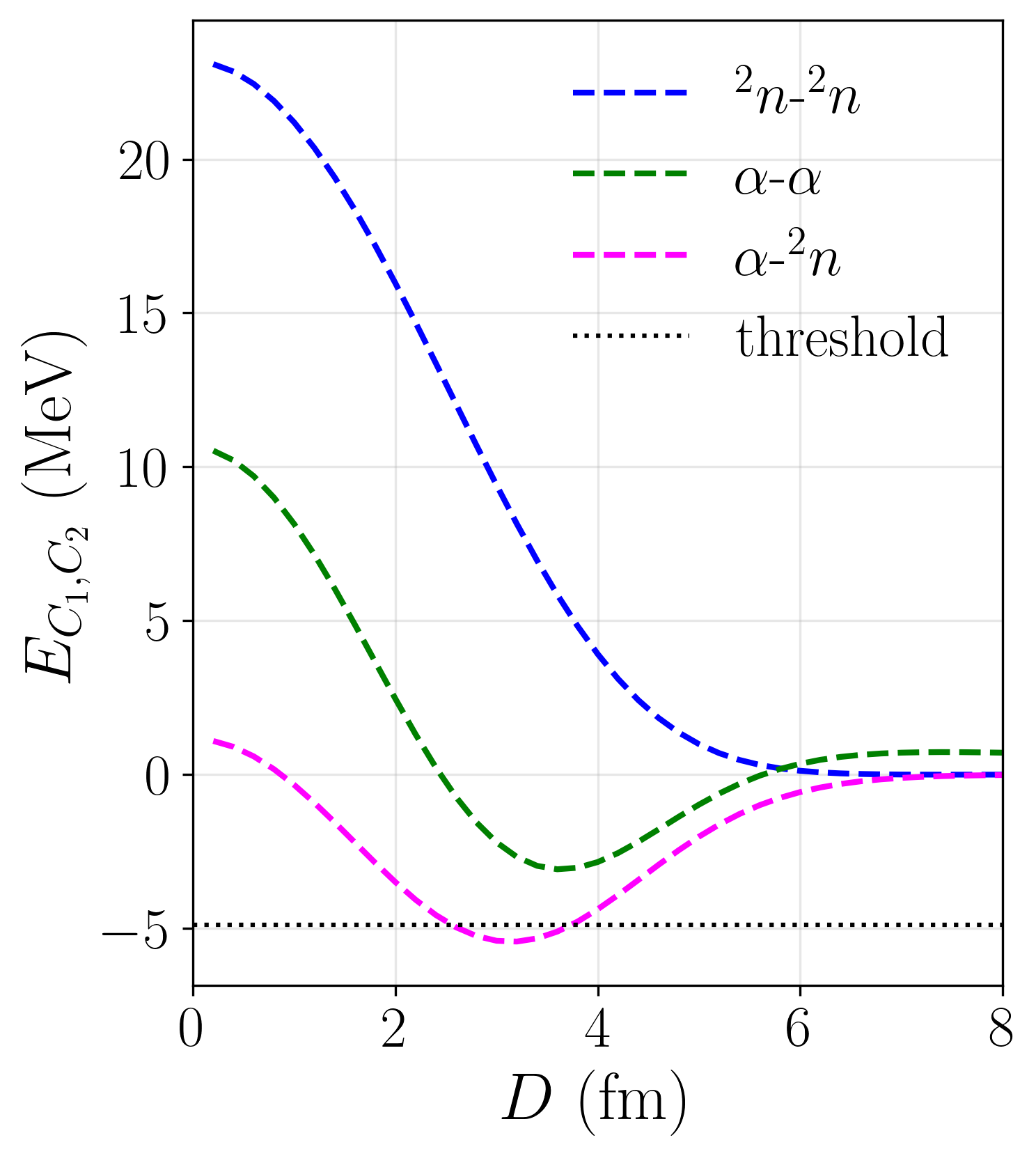}
  \caption{Inter-cluster energies $E_{C_1,C_2}(D)$ for $(C_1,C_2)=({}^2n,{}^2n),(\alpha,\alpha),({}^2n,\alpha)$ with distance $D$ calculated by the original interaction.
  The $\nton$, $\atoa$, and $\aton$ interactions are plotted with the dotted blue, green, and cyan curves, respectively.
  The black dotted line shows the $C_1+C_2$ threshold energy ($-\hbar \omega/4$).}
  \label{fig:PES.tot}
\end{figure}

The energy expectation values of $C_1+C_2$ cluster wave functions in Eq.~\eqref{2body} at the inter-cluster distance $D=|\bm{R}_1-\bm{R}_2|$ for $(C_1,C_2)=({}^2n,{}^2n)$, $({}^2n,\alpha)$, $(\alpha,\alpha)$ is calculated as
\begin{equation}
  E(C_1+C_2;D)=\frac{\braket{C_1+C_2;D|\hat{H} P^{0+}|C_1+C_2;D}}{\braket{C_1+C_2;D|P^{0+}|C_1+C_2;D}}.
\end{equation}
We measure the inter-cluster energy between $C_1$ and $C_2$ with the distance $D$ from the asymptotic energy at $D \to \infty$ as,
\begin{equation}
      E_{C_1,C_2}(D)=E(C_1+C_2;D) - E(C_1+C_2;\infty),
\end{equation}
\begin{equation}
  E(C_1+C_2;\infty) = E(C_1) + E(C_2) + \frac{\hbar \omega}{4},
\end{equation}
where $E(C_i)$ denotes the energy expectation value of a single $C_i$, and $\hbar\omega/4$ is the zero-point energy of the relative motion, \textit{i.e.} the kinetic energy cost to keep cluster center positions at a certain distance.
$E_{C_1,C_2}(D)$ consists of the kinetic part $T_{C_1,C_2}(D)$, the nuclear potential part $U_{C_1,C_2}(D)$ and the Coulomb potential part $U_{C_1,C_2}^{coul}(D)$ as
\begin{equation} \label{eq:decomp}
  E_{C_1,C_2}(D) = T_{C_1,C_2}(D) + U_{C_1,C_2}(D) + U_{C_1,C_2}^{coul}(D).
\end{equation}
Here, $T_{C_1,C_2}(D)$, $U_{C_1,C_2}(D)$, and $U_{C_1,C_2}^{coul}(D)$ are energies of the kinetic, nuclear potential, and Coulomb terms measured from the asymptotic values at $D\to \infty$.
The calculated $E_{C_1,C_2}(D)$ is plotted in Fig.~\ref{fig:PES.tot}.

In all the systems, the inter-cluster energy $E_{C_1,C_2}(D)$ shows short-range repulsion in the small $D$ region due to the energy loss caused by the Pauli blocking effect of nucleons between two clusters.
In the middle region around $D = 3-4$~fm, $\aton$ and $\atoa$ form energy pockets, whereas $\nton$ has no energy pocket.
In the outer region $D > \SI{5.5}{fm}$, $E_{{}^2n,{}^2n}$ and $E_{\alpha,{}^2n}$ approach zero, whereas $E_{\alpha,\alpha}$ is slightly raised by the Coulomb force.
The Coulomb force contribution is relatively small.

\begin{figure}
  \centering
  \includegraphics[width=\linewidth]{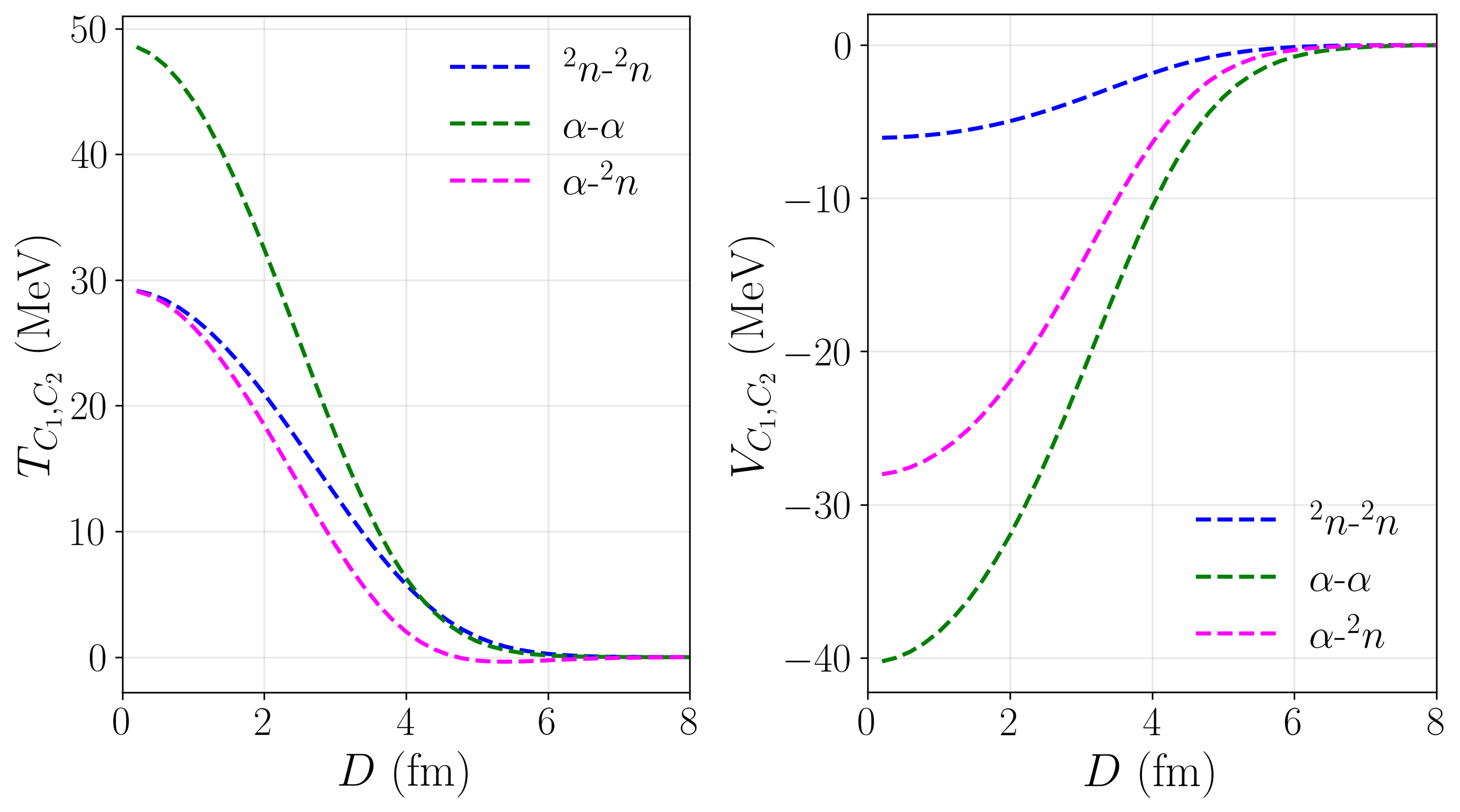}
  \caption{Kinetic part $T_{C_1,C_2}(D)$ (left panel) and nuclear potential part $U_{C_1,C_2}(D)$ (right panel) of inter-cluster energy for $(C_1,C_2)=({}^2n,{}^2n),(\alpha,\alpha),({}^2n,\alpha)$ with distance $D$. $T_{C_1,C_2}$ and $U_{C_1,C_2}$ are measured from the asymptotic values at $D\to\infty$. The colors of the lines correspond to those of Fig.~\ref{fig:PES.tot}.}
  \label{fig:PES.kinpot}
\end{figure}

The short-range repulsion and middle-range behavior of $E_{C_1,C_2}(D)$ are mainly contributed by the kinetic and nuclear potential terms as the Coulomb term gives only minor contributions.
The kinetic part $T_{C_1,C_2}(D)$ and nuclear potential part $U_{C_1,C_2}(D)$ are plotted in the left and right panels of Fig.~\ref{fig:PES.kinpot}, respectively.

The kinetic part $T_{C_1,C_2}(D)$ provides the short-range repulsion, and the nuclear potential part $U_{C_1,C_2}(D)$ gives the middle-range attraction for all three cases.
The nuclear potential parts $U_{C_1,C_2}(D)$ for $\aton$ and $\atoa$ are deep enough to produce the energy pockets of the inter-cluster energy $E_{C_1,C_2}(D)$, whereas that for $\nton$ is shallow and too weak to compensate the kinetic repulsion.
As a result of the balance between the kinetic repulsion and nuclear potential attraction, the energy systematics $E_{{}^2n,{}^2n} > E_{\alpha,\alpha} > E_{\alpha,{}^2n}$ is obtained.
It indicates that the $\alpha+{}^2n+{}^2n$ system is formed under the condition of unbalanced inter-cluster interactions as $E_{{}^2n,{}^2n} \gg E_{\alpha,{}^2n}$, which may suppress the 3-body cluster-gas component in $\He(0_2^+)$.

Let us discuss the energy systematics from the microscopic point of view considering nucleon degrees of freedom.
For the kinetic energy repulsion $T_{C_1,C_2}$ shown in the left panel of Fig.~\ref{fig:PES.kinpot}, one can see the energy systematics $T_{{}^2n,{}^2n} \approx T_{\alpha,{}^2n} < T_{\alpha,\alpha}$ with the ratio $T_{{}^2n,{}^2n}:T_{\alpha,{}^2n}:T_{\alpha,\alpha} \approx 1:1:2$.
The repulsion of the kinetic part $T_{C_1,C_2}$ is caused by single-nucleon kinetic energy loss due to the Pauli blocking between two clusters.
As the second cluster $C_2$ approaches the first cluster $C_1$, $0s$-orbit nucleons in the $C_2$ cluster are blocked by nucleons in the $C_1$ cluster, and finally, at $D\to 0$, they are excited into the $0p$ orbit, which is the lowest allowed orbit free from the Pauli blocking, and the total wave function becomes equivalent to the one-center shell model wave function $(0s)^{A_1}(0p)^{A_2}$~\cite{BBtheorem}.
This Pauli blocking effect provides the monotonically increasing repulsion of the kinetic part in the inter-cluster interaction toward $D\to 0$ with the ratio $2:2:4$ for $\nton :\aton :\atoa$ in proportion to $A_2$ (the nucleon number in the $C_2$ cluster).
As a result, the kinetic energy part at $D\to 0$ before the $0^+$ projection can be simply evaluated by the kinetic energy cost to excite $A_2$ nucleons from $0s$ to $0p$ orbits as $2\hbar\omega$, $2\hbar\omega$, and $4\hbar\omega$ for $\nton$, $\aton$, and $\atoa$, respectively,
and those after the $0^+$ projection are given as
\begin{align}
&T_{C_1,C_2}(D \to 0) \notag \\
&\,\, =  \frac{\hbar \omega}{2} \times A_2 +\frac{\hbar \omega}{2} \label{Tprox1}\\
&\,\, =
  \begin{cases}
    3\hbar \omega/2  & \text{for }(C_1,C_2)=(\alpha,{}^2n), ({}^2n,{}^2n)\\
    5\hbar \omega/2  & \text{for }(C_1,C_2)=(\alpha,\alpha).
  \end{cases}
  \label{Tprox2}
\end{align}
The second term $\hbar\omega/2$ of Eq.~\eqref{Tprox1} comes from the constant energy gain at $D\to\infty$ (no kinetic energy gain at $D\to 0$) by the $0^+$ projection.

For the nuclear potential attraction $U_{C_1,C_2}(D)$ shown in the right panel of Fig.~\ref{fig:PES.kinpot}, the energy systematics $U_{{}^2n,{}^2n} > U_{\alpha,{}^2n} > U_{\alpha,\alpha}$ is obtained.
We consider $NN$ force contributions of $U_{C_1,C_2}(D)$ in the partial wave decomposition into the ${}^1E$, ${}^3E$, ${}^1O$, and ${}^3O$ channels as done in Eq.~\eqref{cent.equiv}.
At $D\to 0$, in which the $C_1+C_2$ cluster wave function goes to the one-center shell model wave function $(0s)^{A_1}(0p)^{A_2}$, the nuclear potential part is expressed as
\begin{equation}  \label{U2cluster}
  \begin{aligned}
    U_{C_1,C_2} &= N_{1E}^{C_1C_2} \langle V({}^{1}E) \rangle_{(0s)(0p)} + N_{3E}^{C_1C_2} \langle V({}^{3}E) \rangle_{(0s)(0p)} \\
    &+ N_{1O}^{C_1C_2} \langle V({}^{1}O) \rangle_{(0s)(0p)} + N_{3O}^{C_1C_2} \langle V({}^{3}O) \rangle_{(0s)(0p)} \\
    &- N_{1E}^{C_2} \bigl[ \langle V({}^{1}E) \rangle_{(0s)^2} - \langle V({}^{1}E) \rangle_{(0p)^2} \bigr]\\
    &- N_{3E}^{C_2} \bigl[ \langle V({}^{3}E) \rangle_{(0s)^2} - \langle V({}^{3}E) \rangle_{(0p)^2} \bigr]
  \end{aligned}
\end{equation}
where $N_{1E}^{C_1C_2}$, $N_{3E}^{C_1C_2}$, $N_{1O}^{C_1C_2}$ and $N_{3O}^{C_1C_2}$ are the numbers of $NN$ pairs between $C_1$ and $C_2$ clusters in the corresponding spin and parity states, and $N_{1E}^{C_2}$ and $N_{3E}^{C_2}$ are those of $NN$ pairs in the $C_2$ cluster.
Here, $\langle V({}^{1}E) \rangle_{(0s)(0p)}$ means the expectation value of the nuclear force for a pair of $0s$ and $0p$ nucleons in the ${}^{1}E$ channel, and likewise for $(0s)^2$ $NN$ pairs and $(0p)^2$ $NN$ pairs.
The numbers $N_{C_i}$ $(C_i={}^{1}E,{}^{3}E,{}^{1}O,{}^{3}O)$ of $NN$ pairs for $(C_1,C_2)=({}^2n,{}^2n)$, $(\alpha,{}^2n)$, $(\alpha,\alpha)$ are listed in Table~\ref{tab:coef2U}.

\begin{table}
  \centering
  \caption{The numbers of the interacting $NN$ pairs for the direct and exchange terms in the inter-cluster potential shown in Eq.~\eqref{U2cluster} for $\nton$, $\aton$, and $\atoa$ pairs.}
  \label{tab:coef2U}
  \begin{ruledtabular}
    \begin{tabular}{cccccccc}
      &$N_{1E}^{C_1C_2}$&$N_{3E}^{C_1C_2}$&$N_{1O}^{C_1C_2}$&$N_{3O}^{C_1C_2}$&$N_{1E}^{C_2}$&$N_{3E}^{C_2}$\\
      \hline
      $\nton$&      1     &     0           &0                &  3              &    1         &   0\\
      $\aton$&     1.5    &     1.5         &       0.5       &  4.5            &    1         &   0\\
      $\atoa$&      3     &     3           &       1         &  9              &    2         &   4\\
      \end{tabular}
  \end{ruledtabular}
\end{table}

Since the even-channel $NN$ forces give dominant attractive contributions and the odd-channel contributions are weak repulsion and minor in nuclear binding, we here consider a rough estimation by ignoring the odd component and the ${}^{1}E$ and ${}^{3}E$ $NN$ force difference as a leading order approximation as
$\langle V({}^{1}E) \rangle \approx \langle V({}^{3}E) \rangle=\langle V(E) \rangle$,
$\langle V({}^{1}O) \rangle \approx \langle V({}^{3}O) \rangle \approx 0$
and obtain
\begin{equation}
  \begin{split}
    U_{C_1,C_2} &\approx  \frac{(A_1-1)A_2}{2} \braket{V(E)}_{(0s)(0p)} \\
        -& \frac{(A_2-1)A_2}{2} \left[ \braket{V(E)}_{(0s)^2} - \braket{V(E)}_{(0p)^2} \right],
  \end{split}
\end{equation}
\begin{equation}
  \begin{split}
    U_{{}^2n,{}^2n} \approx \braket{V(E)}_{(0s)(0p)} - \left[ \braket{V(E)}_{(0s)^2}- \braket{V(E)}_{(0p)^2} \right],
  \end{split}
\end{equation}
\begin{equation}
  \begin{split}
    U_{\alpha,{}^2n}  \approx 3 \braket{V(E)}_{(0s)(0p)} - \left[ \braket{V(E)}_{(0s)^2}- \braket{V(E)}_{(0p)^2} \right],
  \end{split}
\end{equation}
\begin{equation}
  \begin{split}
    U_{\alpha,\alpha}  \approx 6 \braket{V(E)}_{(0s)(0p)} -6 \left[ \braket{V(E)}_{(0s)^2}- \braket{V(E)}_{(0p)^2} \right].
  \end{split}
\end{equation}
The first term is the contribution of $NN$ pairs between $C_1$ and $C_2$ clusters, and the second term corresponds to the internal potential energy loss of the $C_2$ cluster for pairs of two $0p$ nucleons.
The former gives attraction with the ratio $1:3:6$ for $\nton:\aton:\atoa$ as given by pair counting $(A_1-1) A_2 / 2$, which is half (fraction of the even component) of $NN$ pairs of different particles between $C_1$ and $C_2$ clusters.
The latter is a repulsive Pauli effect corresponding to the potential energy loss in the configuration change $(0s)^{A_2}\to(0p)^{A_2}$ of the $C_2$ cluster, and the ratio $1:1:6$ for $\nton:\aton:\atoa$ is obtained by pair counting $(A_2-1) A_2/2$ in $C_2$ cluster.
It should be noted that, in the direct part of the nuclear potential energy $U_{C_1,C_2}$ at the $D\to 0$ limit, the second part vanishes and $U_{C_1,C_2}$ is approximately given by pair counting of the first term as $U_{C_1,C_2}\approx (1/2)A_1 A_2\langle V(E)\rangle_{(0s)(0s)}$, resulting in the ratio $2:4:8$ instead of $1:3:6$.
Therefore, it is naively expected that the inter-cluster interaction between heavier clusters has deeper nuclear potential energy.
As shown in the right panel of Fig.~\ref{fig:PES.kinpot}, the energy hierarchy $|U_{{}^2n,{}^2n}|<|U_{\alpha,{}^2n}|<|U_{\alpha,\alpha}|$ of the calculated result is consistent with this ordering, but the ratio significantly deviates from the naive expectation $2:4:8$; compared to the $\aton$ potential, the $\nton$ potential is shallower than $U_{\alpha,\alpha}/4$ and the $\aton$ potential is deeper than $U_{\alpha,\alpha}/2$ naively expected from the pair counting of the direct part.
The reason for the shallow $\nton$ potential ($U_{{}^2n,{}^2n}$) is a strong Pauli effect of $NN$ pairs of identical particles between ${}^2n$ and ${}^2n$ clusters.
The deep $\aton$ potential ($U_{\alpha,{}^2n}$) can be understood by the repulsion of the second term relatively weaker than that in $U_{\alpha,\alpha}$ between two $\alpha$ clusters, which suffer the strong repulsive effect of the second term due to the large internal energy loss of the $C_2=\alpha$ cluster.

As a result of the kinetic and nuclear potential contributions, the total energy hierarchy $E_{{}^2n,{}^2n}>E_{\alpha,\alpha}>E_{\alpha,{}^2n}$ of the inter-cluster interaction is obtained.
In particular, the Pauli repulsive effects in the kinetic and nuclear potential parts discussed above give significant contributions to the systematics of the inter-cluster interactions $E_{C_1,C_2}$ in the $D=3-4$~fm region; $E_{{}^2n,{}^2n}>0>E_{\alpha,\alpha}>E_{\alpha,{}^2n}$ shown in Fig.~\ref{fig:PES.tot}.
The repulsive $\nton$ interaction $E_{{}^2n,{}^2n}>0$ is described by the strong Pauli effects of $NN$ pairs of two identical particles between clusters.
As for the $\atoa$ and $\aton$ interactions, the energy pocket of $E_{\alpha,\alpha}$ is shallower than $E_{\alpha,{}^2n}$ because of the stronger Pauli effect of the internal potential energy loss of the $C_2=\alpha$ cluster than that of the $C_2={}^2n$ cluster for $\aton$, in addition to the Pauli repulsion of the kinetic part.
It should be stressed again that Pauli effects play crucial roles in both the kinetic and nuclear potential parts, which are microscopic effects from the nucleon antisymmetrization peculiar to interactions between composite clusters consisting of fermions.

\subsection{$\alpha+{}^2n+{}^2n$ cluster-gas state}
In our previous research of $\He$~\cite{nakagawa2025alpha}, we found that
$\He(0_2^+)$ has the dominant 3-body cluster-gas component of $\alpha+{}^2n+{}^2n$ with a significant mixing of the 2-body component of the ($\alpha+{}^2n)+{}^2n$ structure.
This mixing of the second component is expected to be derived from the unbalance between $\aton$ and $\nton$ cluster interactions as $E_{\alpha,{}^2n}(D)<0<E_{{}^2n,{}^2n}(D)$ (see Fig.~\ref{fig:PES.tot}).
Such a much stronger $\aton$ attraction compared to the $\nton$ interactions may induce the $\alpha+{}^2n$ correlation to form a ${}^6\mathrm{He}$-like cluster of $\alpha+{}^2n$ in the $\alpha+{}^2n+{}^2n$ system and favor the 2+1 cluster structure of the 3-cluster system.
Looking at it from the other side, if the $\aton$ and $\nton$ cluster interactions are comparable to each other, the 2-body ($\alpha+{}^2n)+{}^2n$ cluster component could be suppressed.
To confirm the reason for the second component mixing, we perform a test calculation by artificially changing the inter-cluster interactions and demonstrate that
the cluster-gas limit can be produced in a $\alpha+{}^2n+{}^2n$ system with balanced inter-cluster interactions as seen in the $3\alpha$ cluster-gas state of $\C(0_2^+)$.

In order to control the inter-cluster interactions, we tune the original parameters of the central $NN$ forces by hand so as to adjust the $\nton$ and $\aton$ interactions comparable with the $\atoa$ interaction.
For the parameter tuning, we change the strength parameters, $V_{1O}$ and $V_{3O}$, of odd terms in Eq.~\eqref{cent.equiv} and use the original strengths, $V_{1E}$ and $V_{3E}$, for the even terms to keep the internal energy of each cluster unchanged.
Thus tuned parameter set for the balanced $\nton$ and $\aton$ interactions is listed in Table~\ref{tab:intHe} together with the original parameter set.

\begin{table}
  \centering
  \caption{Strength of two nucleon interactions for ${}^1E$, ${}^3E$, ${}^1O$ and ${}^3O$ channels and the corresponding central force parameters in Eq.~\eqref{Volkov} used for the balanced $\He$ calculation.
  Those of the original calculation are also listed.}
  \label{tab:intHe}
  \begin{ruledtabular}
    \begin{tabular}{ccccccccc}
               & $V_{1E}$ & $V_{3E}$ & $V_{1O}$ & $V_{3O}$ & $M$ & $W$ & $B$ & $H$ \\
               \hline
      balanced & 1.0 & 1.0 & -11.92 & 0.88 & 3.26 & -2.26 & 3.20 &-3.20 \\
      original & 1.0 & 1.0 & -0.16 & -0.16 & 0.58 & 0.42 & 0.0 & 0.0
    \end{tabular}
  \end{ruledtabular}
\end{table}

\begin{figure}
  \includegraphics[width=\linewidth]{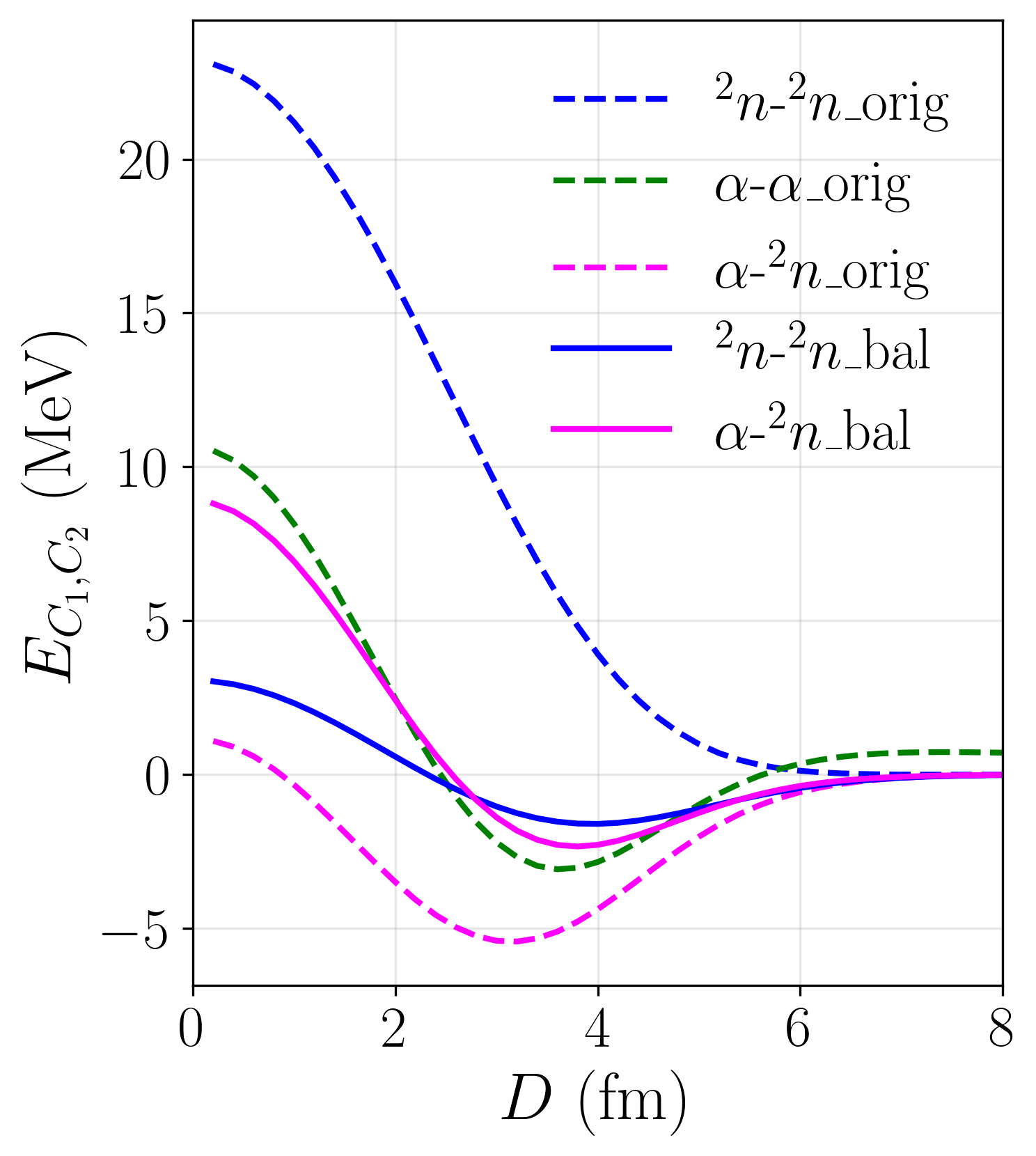}
  \caption{Inter-cluster energies $E_{C_1,C_2}(D)$ of $\nton$ and $\aton$ obtained by the original and balanced parameter sets of the central $NN$ forces given in Table~\ref{tab:intHe}.
  The results of $\nton$ and $\aton$ for the balanced parameter set are shown by the solid blue and cyan curves, respectively.
  The original results of $\nton$, $\atoa$, and $\aton$ energies are plotted with blue, green and cyan dotted lines, respectively.}
  \label{fig:PES.8he}
\end{figure}

After the interaction changing, the $\nton$ and $\aton$ interactions are actually balanced, and in agreement with the original $\atoa$ interaction as shown in Fig.~\ref{fig:PES.8he}.

Using the tuned parameter set of the central $NN$ forces for the balanced inter-cluster interactions, we recalculate $\He$ wave functions of Eq.~\eqref{wf8he} in the same framework as done in the previous work.
It should be commented that the default $\He$ calculation using the original parameter set of the $NN$ force (called original $\He$ calculation in this paper) reproduces properties of the physical $\He$ such as the ground-state energy, radius, and the $0_2^+$ excitation energy as well as the ${}^2n$ separation energies of $\He$ and ${}^6\mathrm{He}$.
The original set also reproduces the $\atoa$ scattering phase shift.
The calculation using the tuned parameter set of the $NN$ force for the balanced inter-cluster interactions, which we call "balanced calculation", corresponds not to the physical $\He$ but to an unphysical "$\He$" artificially created to examine how clusters behave in the $0_2^+$ state under the condition of the balanced inter-cluster interactions.

Table~\ref{tab:balHe} shows results of the binding energy (B.E.), the $0_2^+$ excited energy $E_x(0_2^+)$, root-mean-squared (r.m.s.) radii $r_m$ of the $0_{1,2}^+$ states and the IS0 transition matrix element $\mathcal{M}(\mathrm{IS0};0_1^+ \to 0_2^+)$ obtained in the balanced case of the $\He$ calculation and also those of the original (default) $\He$ calculation taken from Ref.~\cite{nakagawa2025alpha}.

\begin{table}
  \centering
  \caption{The binding energies (B.E.), $0_2^+$ excitation energies, r.m.s. radii of the $0_{1,2}^+$ states, and the IS0 transition matrix elements $\mathcal{M}(\mathrm{IS0};0_1^+ \to 0_2^+)$ of the balanced $\He$.
  The results of original $\He$ and $\C$~\cite{nakagawa2025alpha} are also listed.}
  \label{tab:balHe}
  \begin{ruledtabular}
    \begin{tabular}{ccccccc}
           & B.E. &$E_x(0_2^+)$ & $r_{m}(0_1^+)$ & $r_{m}(0_2^+)$ & $\mathcal{M}(\mathrm{IS0})$ \\
           & (MeV) & (MeV) & (fm) & (fm) & (\si{fm^2}) \\
           \hline
           balanced $\He$ & 33.79 & 10.43 & 2.08 & 3.30 & 7.15 \\
           original $\He$ & 29.44 & 5.20 & 2.40 & 3.89 & 15.5 \\
           original $\C$  & 90.02 & 8.04 & 2.38 & 3.18 & 8.08 \\
    \end{tabular}
  \end{ruledtabular}
\end{table}

In the balanced $\He$ calculation, the $0_2^+$ energy $-\SI{23.36}{MeV}$ is obtained at almost the same energy as $-\SI{24.24}{MeV}$ of the original calculation and near the $\alpha+{}^2n+{}^2n$ energy threshold as well.
The larger radii of the $0_2^+$ state than the ground state in the balanced calculation are due to the $\alpha+{}^2n+{}^2n$ cluster structure. Compared to the original (default) $\He$ calculation, the balanced calculation obtains r.m.s. radii and $\mathcal{M}(\mathrm{IS0})$ slightly smaller than the default calculation showing a size shrinkage from the original $\He$ wave functions.
Actually, the values of $r_m(0_2^+)$ and $\mathcal{M}(\mathrm{IS0})$ of the balanced $\He$ are similar to those of the $\C$ wave function as expected from the fact that the $\nton$ and $\aton$ interactions in the balanced case are controlled to reproduce the $\atoa$ interactions.

\begin{figure}
  \centering
  \includegraphics[width=\linewidth]{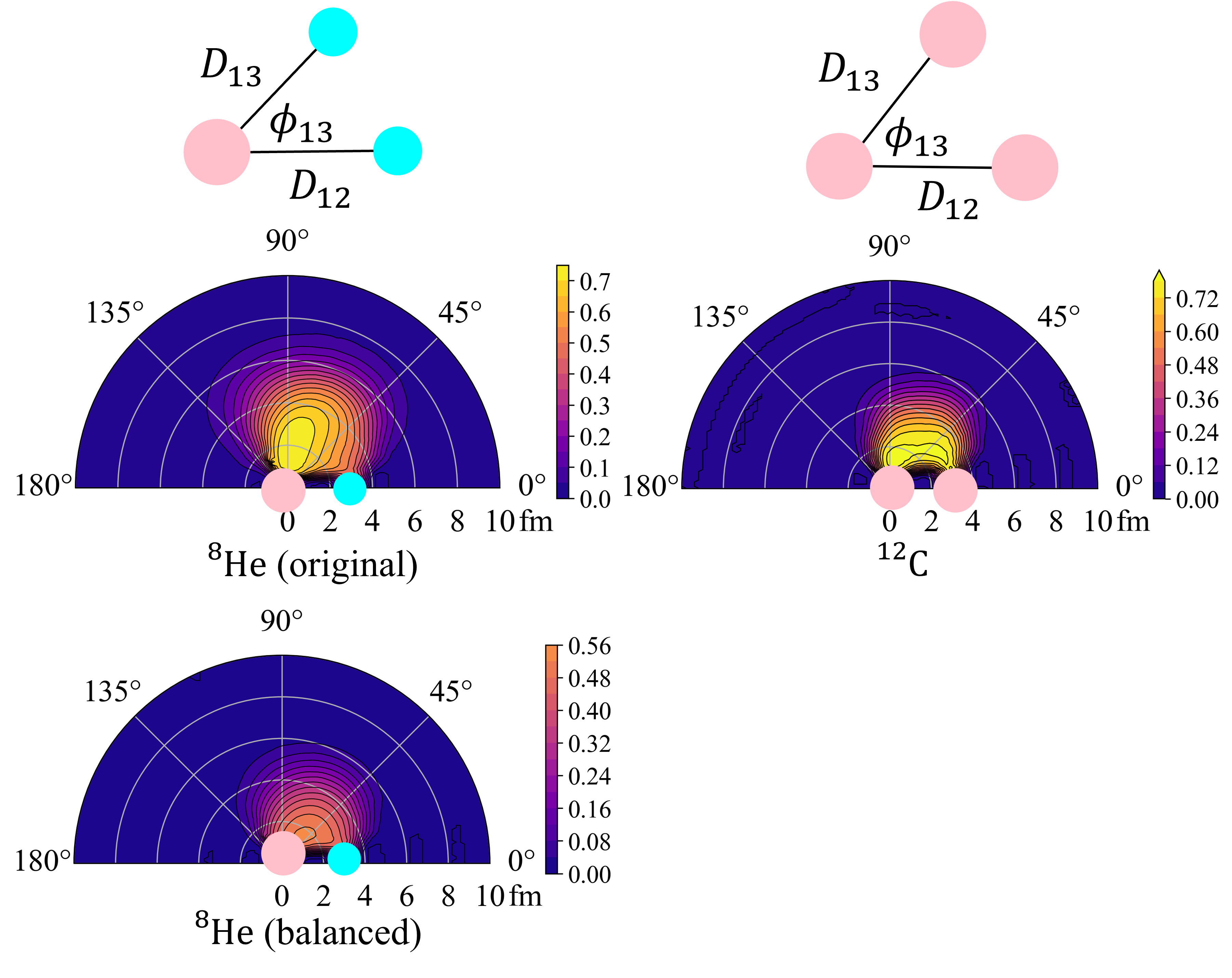}
  \caption{Spatial distributions of the $C_3$ cluster in the $0_1^+$ of the balanced $\He$, original $\He$ and $\C$.
  The squared overlap ${\cal O}^{C_1+C_2+C_3}(\bm{R_1},\bm{R_2},\bm{R_3})$ given in Eq.~\eqref{ougi} is plotted on the $(D_{13},\phi_{13})$ plane around the $C_1$ and $C_2$ clusters located on the horizontal axis at $(0,0)$ and $(D_{12}=\SI{3}{fm},0)$, respectively.
  For the balanced $\He(0_1^+)$ (bottom-left) and the original $\He(0_1^+)$ (middle-left) case, $(C_1,C_2,C_3)=(\alpha,{}^2n,{}^2n)$ is adopted, and for the $\C$ case (middle-right), $C_1=C_2=C_3=\alpha$ is adopted.
  The center position of the fixed $\alpha$ clusters and the dineutron clusters are drawn with pink and light blue circles for the balanced and original $\He(0_1^+)$ plots, and those of the fixed $2\alpha$ clusters are plotted with pink circles on the $\C$ plot as well.}
  \label{fig:BO8he1}
\end{figure}

We briefly describe the ground state structure, which gives important contribution to the structure of the $0^+_2$ state through the orthogonality.
We calculate the squared overlap of the $\alpha+{}^2n+{}^2n$ cluster configuration given by the BB wave function in Eq.~\eqref{3body} with the $\He(0_1^+)$ wave functions as
\begin{equation}  \label{ougi}
  \begin{split}
    &{\cal O}^{\alpha+{}^2n+{}^2n}(\bm{R_1},\bm{R_2},\bm{R_3}) \\
    &\equiv \left| \braket{\alpha+{}^2n+{}^2n;D_{12},D_{13},\phi_{13}| \He(0_1^+)} \right|^2,
  \end{split}
\end{equation}
where $D_{12} \equiv |\bm{R}_2-\bm{R}_1|$ and $D_{13} \equiv |\bm{R}_3-\bm{R}_1|$ are distances of two ${}^2n$ cluster centers measured from the $\alpha$ cluster, and $\phi_{13}$ is the opening angle between them.
Figure~\ref{fig:BO8he1} shows the spatial distributions of the squared overlaps of the original and balanced $\He(0_1^+)$.

These plots display spatial distributions of one dineutron ($C_3={}^2n$) cluster on the $({D}_{13}, \phi_{13})$ plane around the $C_1=\alpha$ and $C_2={}^2n$ clusters located on the horizontal axis at $(0,0)$ and $(D_{12}=\SI{3}{fm},0)$, respectively (see the schematic figure for the 3-cluster positions in Fig.~\ref{fig:BO8he1}).
We also show the squared overlap for $3\alpha$ cluster configurations of $\C(0^+_1)$ for comparison.
The overlaps are distributed in a region around $\phi_{13} = 0 -\ang{90}$ and $D_{13} = 1 - 4$~fm having a peak around $(\SI{2}{fm}, \ang{73})$.
These results indicate the ground state of $\He$ has a shell-model structure and compact $\alpha+{}^2n+{}^2n$ cluster components.
Also in the case of $\C$, the overlaps distributed in a small distance region between clusters indicate a compact $3\alpha$ structure in the $\C(0^+_1)$.

\begin{figure*}
  \includegraphics[width=12cm]{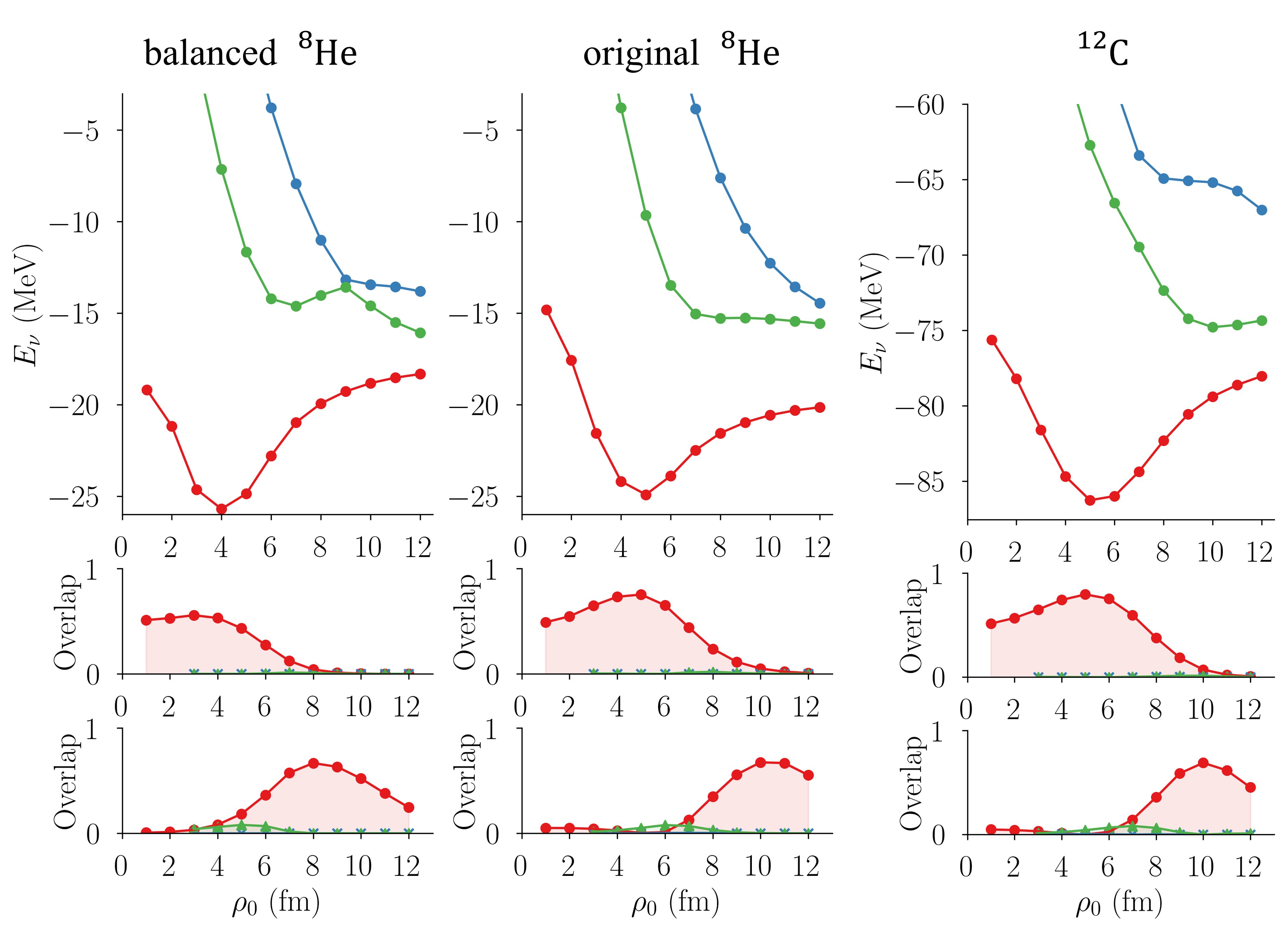}
  \caption{The energy spectra $E_{\nu}(\rho_0)$ for $\nu=1,2,3$ and the $0^+_{\nu}(\rho_0)$ components contained in the $\He(0^+_{1,2})$ wave functions for the balanced calculation compared with the original $\He$ and $\C$ calculations.
  (Top): Energy spectra $E_{\nu}(\rho_0)$ of the $\rho$-fixed 3-body states of the balanced $\He$, original $\He$, and $\C$ in the left, middle, and right panels, respectively.
  (Middle): The $0^+_{\nu}(\rho_0)$ components contained in the $0_1^+$ wave functions of balanced $\He$, original $\He$, and $\C$.
  (Bottom): Those in the $0_2^+$ wave functions.
  The results for $\nu=1$, 2 and 3 are plotted with red, green, and blue colors, respectively.}
  \label{fig:adbt.comp}
\end{figure*}

In order to discuss radial behaviors of the cluster structures in the monopole excitation mode in the balanced $\He$ calculation, we perform the $\rho$-fixed analysis as done in the previous work~\cite{nakagawa2025alpha}.
We prepare $\rho$-fixed states by the truncated GCM calculation within the subspace of $\alpha+{}^2n+{}^2n$ fixed at the hyper radius $\rho=\rho_0$ expressed in Eq.~(29) of Ref.~\cite{nakagawa2025alpha}, and construct the $\nu$-th $0^+$ states at each $\rho_0$.
Fig.~\ref{fig:adbt.comp} shows the $\rho$-fixed results of the energy spectra $E_\nu(\rho_0)$ for $\nu = 1, 2, 3$ and the $0^+_\nu(\rho_0)$ components contained in the $\He(0^+_{1,2})$ wave functions for the balanced calculation compared with the original $\He$ and $\C$ calculations.

The $0^+_\nu$ energy curves along $\rho_0$ in the balanced calculation show behaviors similar to the original one; the energy minimum in the $\rho=4-5$~fm region for the $\nu=1$ state, and a large energy gap between the $\nu=1$ and $\nu=2$ states.
Moreover, the dominance of the $\nu=1$ component in the $\He(0_2^+)$ wave function is common in the balanced and original calculations, indicating that the radial excitation is the primary excitation mode in the balanced $\He(0_2^+)$ state as found in the original $\He(0_2^+)$.

\begin{figure}
  \centering
  \includegraphics[width=\linewidth]{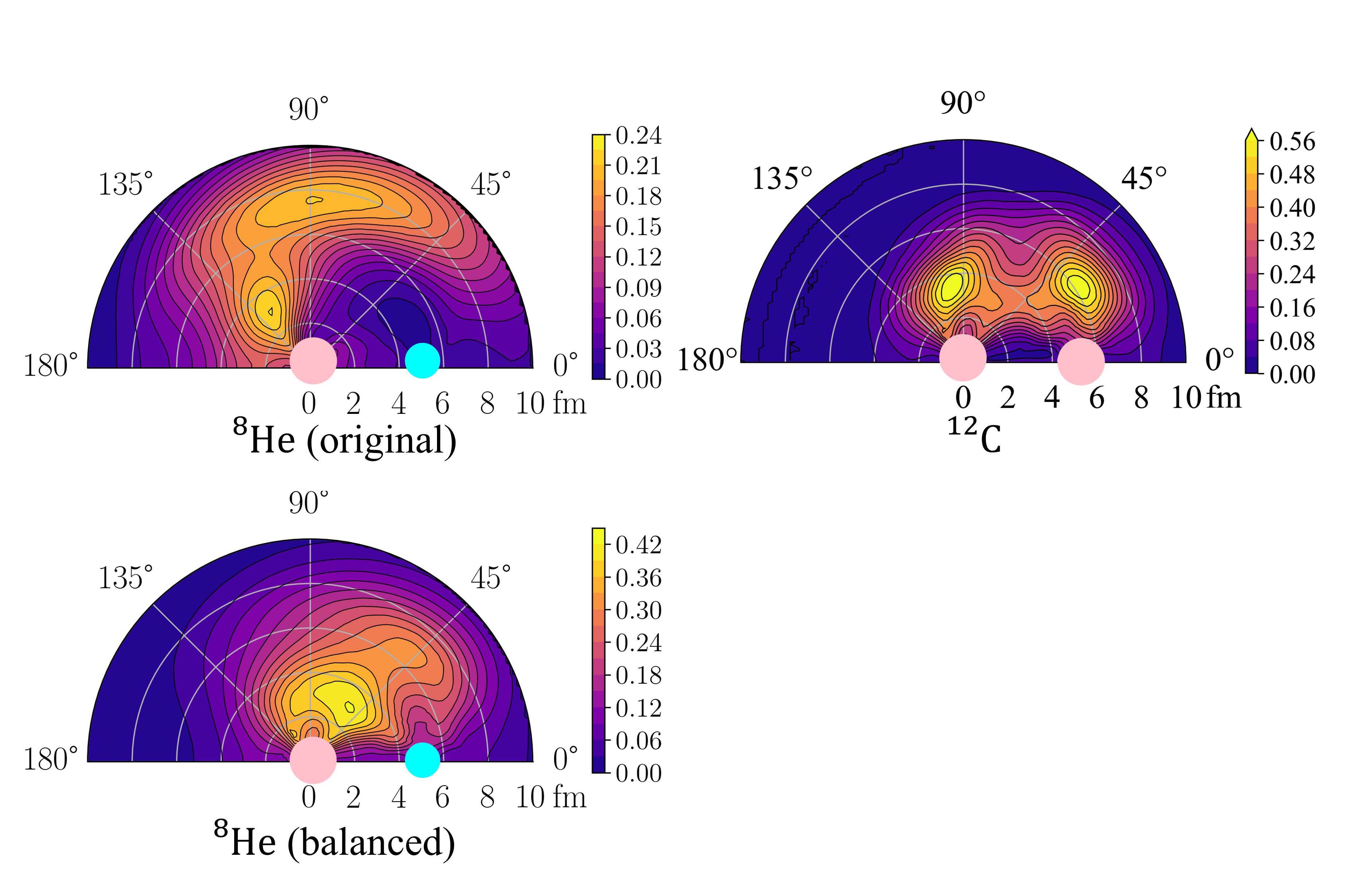}
  \caption{Same as Fig.~\ref{fig:BO8he1} but for the $0_2^+$ states of the balanced $\He$, original $\He$ and $\C$ with $C_2$ at $D_{12}=\SI{5}{fm}$.}
  \label{fig:BO.bal}
\end{figure}

Even though the radial behaviors of the balanced $\He(0_2^+)$ are similar to the original $\He(0_2^+)$ results, it does not necessarily mean that the mixing of the 3-body cluster-gas and 2-body ($\alpha+{}^2n)+{}^2n$ cluster structures is unchanged because these two kinds of cluster components can be inclusively contained in the radial excitation mode in the $\rho$-fixed analysis.
Indeed, the two-component mixing in $\He(0_2^+)$ is changed before and after the interaction changing.
To discuss contributions from two kinds of cluster components, we investigate spatial structures of the $\alpha+{}^2n+{}^2n$ configurations in detail.
Fig.~\ref{fig:BO.bal} shows the squared overlaps for the $0_2^+$ state obtained by the balanced $\He$ calculation together with the original $\He$ and $\C$ results.

One dineutron distribution of the $\alpha+{}^2n+{}^2n$ configurations at $D_{12}=5$~fm in $\He(0_2^+)$ for the balanced and original cases is shown in the left top and bottom panels, respectively, and one $\alpha$ distribution of the $3\alpha$ configurations in the original $\C(0_2^+)$ is shown in the right panel.
As shown in the left bottom, two peaks of the dineutron distribution can be seen in the original $\He(0_2^+)$.
The broad peak in the $D_{13}\sim 7$~fm region indicates the 3-body cluster-gas component, and the sharp peak at $\bm{D}_{13}=(D_{13},\phi_{13}) \sim (\SI{3}{fm},\ang{120})$ corresponds to the 2-body ($\alpha+{}^2n)+{}^2n$ cluster component.
On the other hand, $\He(0_2^+)$ in the balanced case does not show such a two-peak structure of the dineutron distribution but exhibits properties of the 3-cluster gas state.
As shown in the left top panel of Fig.~\ref{fig:BO.bal} for the balanced case, the peak at $\bm{D}_{13}\sim (\SI{3}{fm},\ang{120})$ for the 2-body ($\alpha+{}^2n)+{}^2n$ cluster component vanishes.
Instead, the dineutron amplitude broadly distributes in wide regions on the $\bm{D}_{13}$ plane because two dineutrons are rather freely moving in $S$ wave around the $\alpha$ cluster like a cluster-gas, similarly to the $3\alpha$ cluster-gas state of the original $\C(0_2^+)$.
We will give further discussion on the spatial distribution of two dineutrons later in Section~\ref{sec:dis}.
It should be commented that the dineutron distribution in the balanced $\He(0_2^+)$ slightly shifts inward from that in the original $\He(0_2^+)$ because of the downsizing effect of the $\alpha+{}^2n+{}^2n$ cluster system in the balanced $\He(0_2^+)$,
which has the r.m.s. radius comparable to that of $\C$ but smaller than that of the original $\He(0_2^+)$ as shown in Table~\ref{tab:balHe}.
From the comparison of two calculations before and after the interaction changing, we can say that the 2-body ($\alpha+{}^2n)+{}^2n$ cluster component is suppressed while the 3-body $\alpha+{}^2n+{}^2n$ cluster-gas component is enhanced in $\He(0_2^+)$ after the interaction changing for the balanced calculation.
It is concluded that the 3-cluster gas state in $\He(0_2^+)$ is derived from the balanced inter-cluster interactions, in other words, the ($\alpha+{}^2n)+{}^2n$ structure mixing in the original $\He(0_2^+)$ is induced by the unbalance of the inter-cluster interactions between $\aton$ and $\nton$.

\subsection{clustering of $\Be$} \label{sec:clus10Be}
Similar analysis is performed for $\Be$ by controlling the inter-cluster interactions to discuss the $2\alpha+{}^2n$ structure of $\Be(0_2^+)$.
As a test calculation, we tune again the central $NN$ force so that $\atoa$ and $\aton$ inter-cluster interactions are balanced, and examine whether a 3-body $2\alpha+{}^2n$ cluster-gas state can be produced in the balanced $\Be$ calculation after the interaction changing.
We artificially change the strength parameters, $V_{1O}$ and $V_{3O}$, for odd terms of the central $NN$ force in Eq.~\eqref{cent.equiv} while keeping even terms unchanged from the original strengths.
It is difficult to independently control the $2\alpha$-${}^2n$ and $\atoa$ interactions by changing odd terms because the ratios of $NN$ pair counting $N_{1O}$ and $N_{3O}$ are common between $2\alpha$-${}^2n$ and $\atoa$ interactions.
To solve this problem, we introduce isospin dependence of the $NN$ force by adding an attractive ${}^{3}O$ term $V_{3O}^{pp}P({}^{3}O)$ only for proton-proton pairs, where $P_{pp}({}^{3}O)$ is a projection operator onto the $T_z=1$ channels of the ${}^{3}O$ state.
A new parameter set tuned for the balanced $\Be$ calculation is listed in Table~\ref{tab:intBe}.

\begin{table}
  \centering
  \caption{Strength of two nucleon interactions for ${}^1E$, ${}^3E$, ${}^1O$ and ${}^3O$ channels and that of the additional ${}^3O$ term of the $pp$ channel used for the balanced and original $\Be$ calculations.
  The corresponding central force parameters in Eq.~\eqref{Volkov} are also listed.}
  \label{tab:intBe}
  \begin{ruledtabular}
    \begin{tabular}{ccccccccccc}
               & $^1E$ & $^3E$ & $^1O$ & $^3O$ & $M$ & $W$ & $B$ & $H$ & $V_{3O}^{pp}$ \\
               \hline
      balanced & 1.0 & 1.0 & -0.20 & -0.32 & 0.63 & 0.37 & -0.03 & 0.03 & 0.928\\
      original & 1.0 & 1.0 & -0.20 & -0.20 & 0.60 & 0.40 & 0.0 & 0.0 & 0.0
    \end{tabular}
  \end{ruledtabular}
\end{table}

\begin{figure}
  \includegraphics[width=\linewidth]{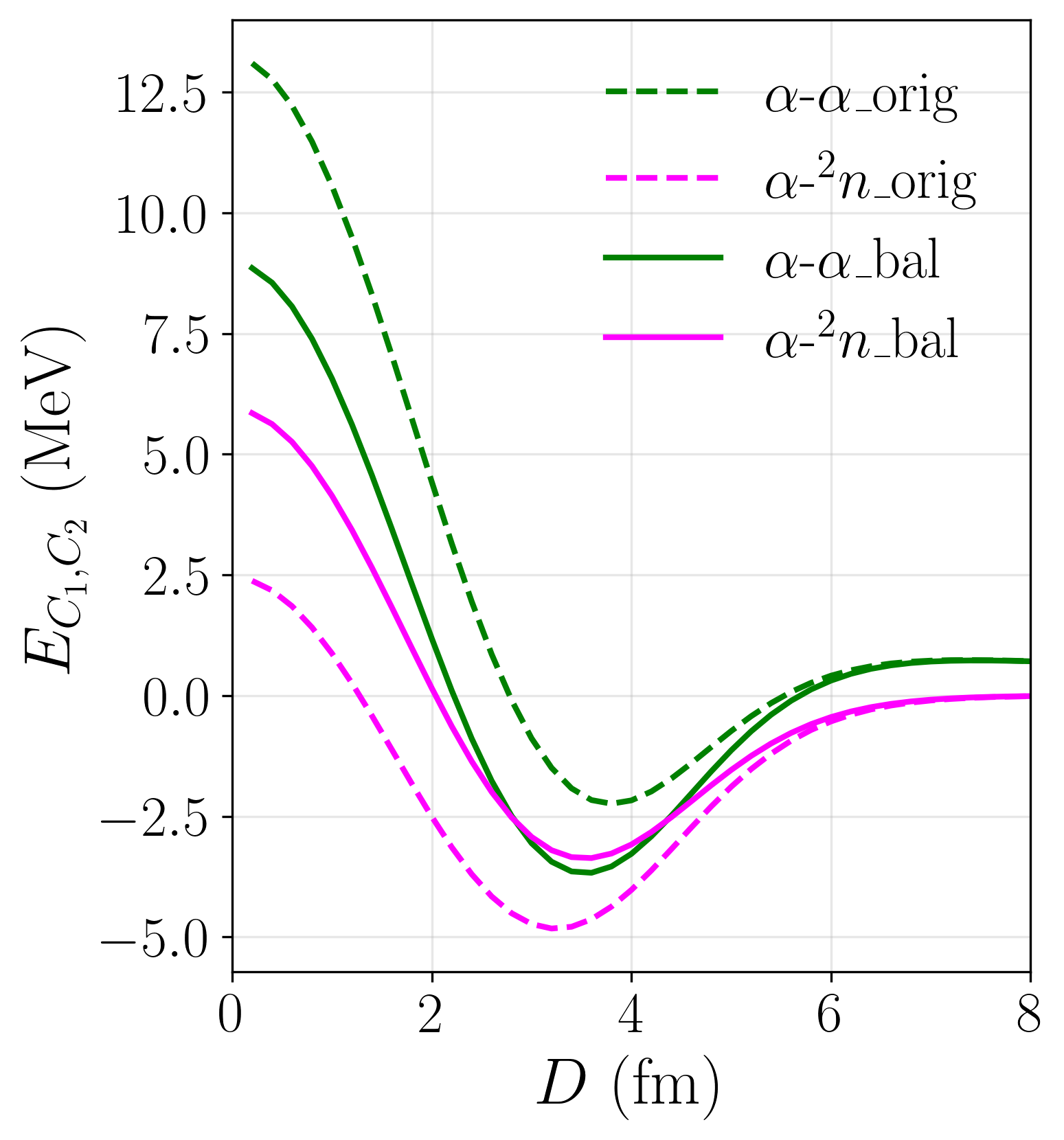}
  \caption{Balanced inter-cluster interactions of $\atoa$ and $\aton$.
  The balanced $\atoa$ and $\aton$ interactions are shown by the solid green and cyan curve, respectively.
  The original $\atoa$ and $\aton$ interactions are also plotted with green and pink dotted curves for comparison.}
  \label{fig:PES.10Be}
\end{figure}

As shown in Fig.~\ref{fig:PES.10Be}, it actually produces the balanced $\aton$ and $\atoa$ interactions, which agree with the original $\atoa$ interaction.

Using the $NN$ force parameter set after the tuning for the balanced inter-cluster interactions, we recalculate $\Be$ in the same framework as the original $\Be$ calculation and obtain results of the balanced calculation.

The balanced $\Be$ calculation approximately reproduces energies and r.m.s. radii of the original results for $\Be(0_1^+)$ and $\Be(0_2^+)$ as shown in Table~\ref{tab:balBe}, in which results of the original $\C$ calculation are also listed as a reference for the cluster-gas limit.

\begin{table}
  \centering
  \caption{The B.E., the $0_2^+$ excited energies, and r.m.s. radii of the $0_{1,2}^+$ states of the balanced and original $\Be$.
  Those of $\C$~\cite{nakagawa2025alpha} are also listed.}
  \label{tab:balBe}
  \begin{ruledtabular}
    \begin{tabular}{cccccc}
           & $E(0_1^+)$ &$E_x(0_2^+)$ & $r_{m}(0_1^+)$ & $r_{m}(0_2^+)$ \\
           & (MeV) & (MeV) & (fm) & (fm) \\
           \hline
           balanced $\Be$ & -60.65 & 6.43 & 2.51 & 3.34 \\
           original $\Be$ & -60.58 & 5.89 & 2.52 & 3.33 \\
                     $\C$ & -90.02 & 8.04 & 2.38 & 3.18 \\
    \end{tabular}
  \end{ruledtabular}
\end{table}

\begin{figure}
  \centering
  \includegraphics[width=\linewidth]{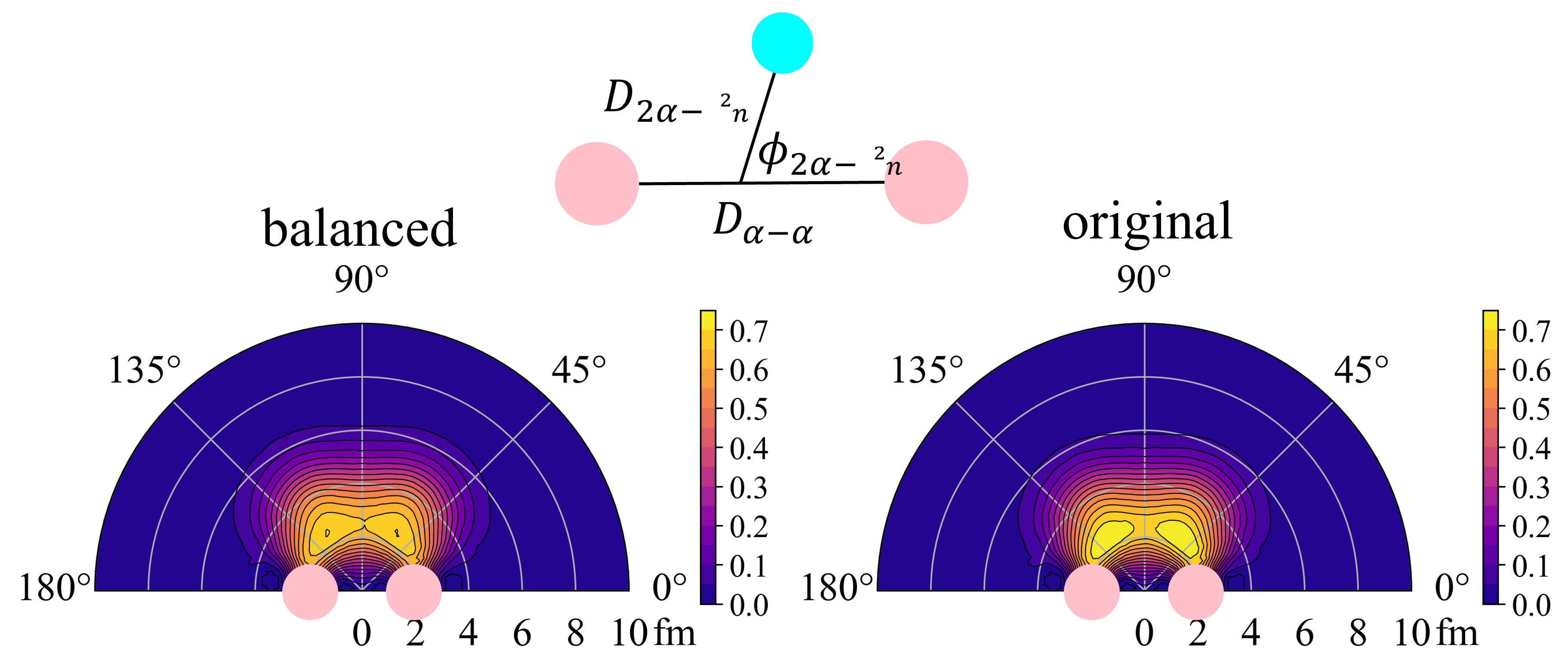}
  \caption{${}^2n$ distribution around the $2\alpha$ clusters of the balanced and original $\Be(0_1^+)$.
  The squared overlap ${\cal O}^{2\alpha+{}^2n}(\bm{R}_1,\bm{R}_2,\bm{R}_3)$ is plotted on the $\bm{D}_{2\alpha\text{-} {}^2n}$ plane, where the $2\alpha$ clusters are located on the horizontal axis with the distance $D_{\atoa}=4$~fm.
  The results of the balanced (original) $\Be$ calculation are shown on the left (right) column.}
  \label{fig:ougi.10Be1}
\end{figure}

For detailed discussions of the cluster structure in $\Be$, we calculate the squared overlaps of the $2\alpha + {}^2n$ cluster BB wave functions with the $\Be$ wave functions obtained by the original and balanced calculations, as
\begin{equation}
  \begin{split}
    &\quad {\cal O}^{2\alpha+{}^2n}(\bm{R}_1,\bm{R}_2,\bm{R}_3)  \\
    &=\left|\braket{2\alpha+{}^2n;D_{\atoa}, D_{2\alpha \text{-} {}^2n}, \phi_{2\alpha \text{-} {}^2n} |\Be} \right|^2,
  \end{split}
\end{equation}
where $D_{\atoa} \equiv |\bm{R}_1-\bm{R}_2|$, $D_{2\alpha \text{-} {}^2n} \equiv |\frac{1}{2}(\bm{R}_1+\bm{R}_2)-\bm{R}_3|$, and $\phi_{2\alpha \text{-} {}^2n}$ is an angle between them.
Fig.~\ref{fig:ougi.10Be1} shows spatial distributions of ${\cal O}^{2\alpha+{}^2n}(\bm{R}_1,\bm{R}_2,\bm{R}_3)$ on $(D_{2\alpha \text{-} {}^2n}, \phi_{2\alpha \text{-} {}^2n})$ plane with fixing $D_{\atoa}=\SI{4}{fm}$ of the balanced and original $\Be(0_1^+)$.

In the original $\Be(0_1^+)$, the ${}^2n$ cluster has concentrated peaks along $D_{2\alpha \text{-} {}^2n}=\SI{2}{fm}$ regions.
This shows a strong spatial correlation between $\alpha$ and ${}^2n$ clusters and $2\alpha+{}^2n$ forms a compact isosceles triangle.

\begin{figure}
  \centering
  \includegraphics[width=\linewidth]{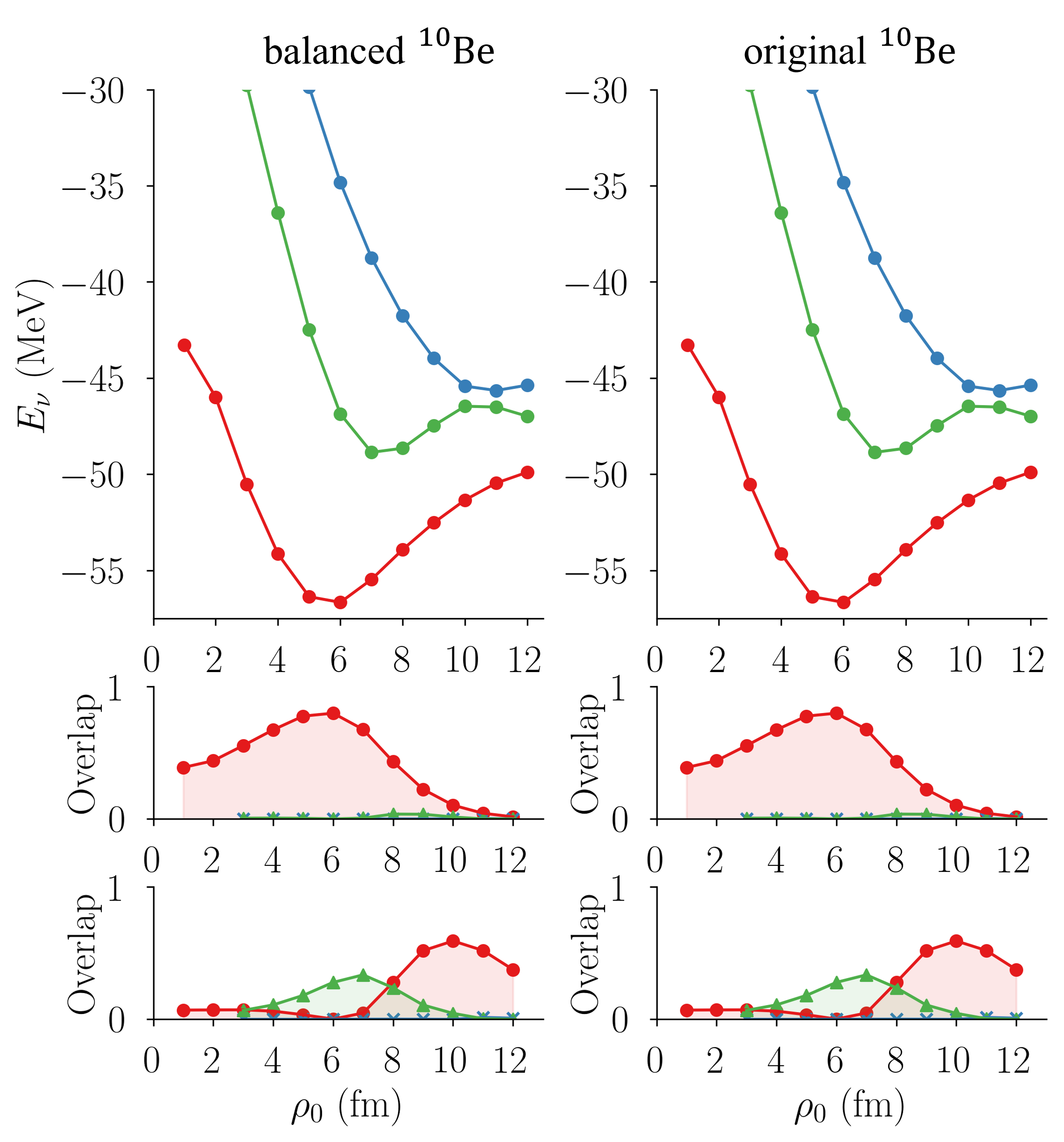}
  \caption{Same as Fig.~\ref{fig:adbt.comp} in the case of the original (left) and balanced (right) $\Be$.}
  \label{fig:adbt.10Be}
\end{figure}

In order to discuss radial behavior of the cluster structures in $\Be(0_2^+)$, we perform the $\rho$-fixed analysis as done for $\He(0_2^+)$ in the previous section.
The $\rho$-fixed results of the original and balanced $\Be$ calculations are compared in Fig.~\ref{fig:adbt.10Be}; the energy spectra $E_{\nu}(\rho_0)$ for $\nu=1,2,3$ and the $0^+_{\nu}(\rho_0)$ components contained in the $\Be(0^+_{1,2})$ wave functions are plotted as functions of the fixed value $\rho=\rho_0$ of the hyperradius parameter.

The two calculations, original and balanced $\Be$, exhibit quite similar results for the $0^+_{\nu}$ energy curves and the $0^+_{\nu}(\rho)$ components in $\Be(0_1^+)$ and $\Be(0_2^+)$.
In both calculations, the $\Be(0_1^+)$ state is dominated by $\nu=1$ components around $\rho=4-7$~fm because of the compact $2\alpha+{}^2n$ triangle structure, whereas the $\Be(0_2^+)$ state contains significant mixing of the $\nu=2$ components in the middle region of $\rho \sim 7$~fm in addition to the $\nu=1$ component around $\rho \sim 10$~fm.
This $\nu=2$ mixing indicates that $\Be(0_2^+)$ is not the radial excitation mode but is induced by the $\nu=1\to\nu=2$ excitation corresponding to a bending mode of three clusters.
This feature is commonly obtained in two calculations, the original and balanced ones, and implies that a 3-body $2\alpha+{}^2n$ cluster-gas state is not created in $\Be(0_2^+)$ even after the interaction changing for the balanced $\atoa$ and $\aton$ interactions.
The $\nu=2$ mixing in $\Be(0_2^+)$ is due to the energy minimum of the $\nu=2$ states.
As shown in the top panels of Fig.~\ref{fig:adbt.10Be}, the $\nu=2$ energy curve has the minimum around $\rho \sim 7$~fm, and hence the $\nu=1\to\nu=2$ excitation is favored and pulled down to a low energy below (or comparable to) the radial excitation along the $\nu=1$ energy curve.
The $\nu=2$ energy minimum still remains in the balanced $\Be$ calculation.
It is in contrast to the $\He(0_2^+)$ and $\C(0_2^+)$ results, which exhibit the radial excitation mode built on the $\nu=1$ energy curve along $\rho$.

\begin{figure}
  \centering
  \includegraphics[width=\linewidth]{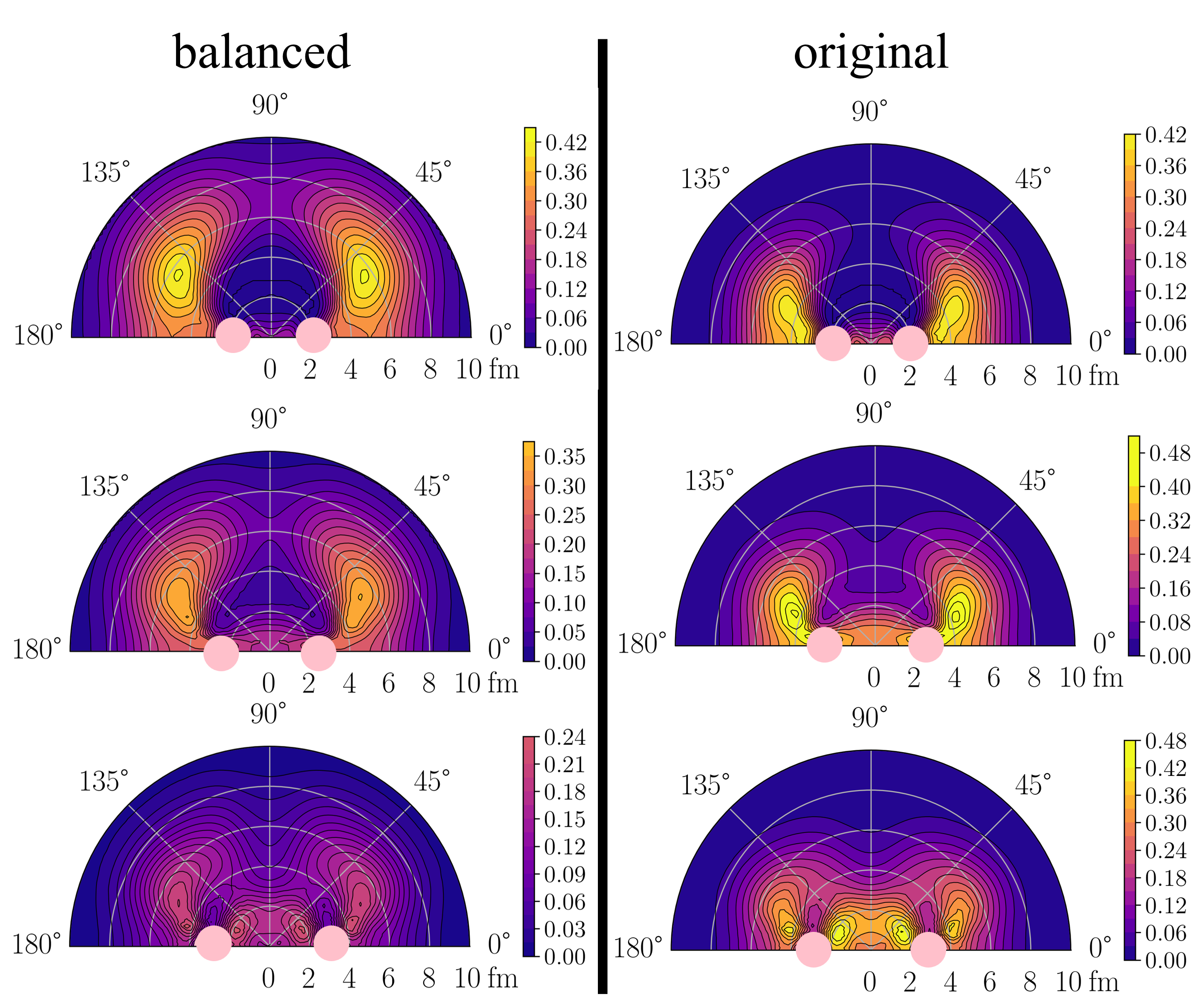}
  \caption{Same as Fig.~\ref{fig:ougi.10Be1} but for the $\Be(0_2^+)$ states.
  The results with $D_{\atoa}=4$, $5$, $6$~fm are shown on the top, middle and bottom rows, respectively.}
  \label{fig:ougi.10Be2}
\end{figure}

The balanced (original) $\Be(0_2^+)$ results of spatial distributions of ${\cal O}^{2\alpha+{}^2n}(\bm{R}_1,\bm{R}_2,\bm{R}_3)$ are shown in the left (right) panels of Fig.~\ref{fig:ougi.10Be2}.

In the original $\Be(0_2^+)$, the overlaps show the dineutron distribution elongated along the $\atoa$ direction;
the stretched $2\alpha+{}^2n$ configurations around small $\phi_{2\alpha\text{-}{}^2n}$ contribute, but the triangle configurations around $\phi_{2\alpha\text{-}{}^2n}\sim 90^\circ$ are suppressed.
It shows a feature of the angular excitation having a nodal behavior on $\phi_{2\alpha\text{-}{}^2n}$, and is associated with the bending mode (or stretching mode) built on the triangle $2\alpha+{}^2n$ configuration.
We comment that the stretched $2\alpha+{}^2n$ configuration in the present result of $\Be(0_2^+)$ corresponds to the results of other theoretical works of $\Be$ in the molecular orbital picture~\cite{ItagakiMO1,ItagakiMO2,ITO2006293,PhysRevC.52.628,PhysRevC.60.064304}.

\section{discussion} \label{sec:dis}
\subsection{inter-cluster distance in $\He$}
\subsubsection{definition of inter-cluster distance distribution}
For further discussions of cluster structures of $\He(0_1^+)$ and $\He(0_2^+)$, we analyze the inter-cluster correlations of $\aton$ and $\nton$ in the $\He$ wave functions.
Here, we consider inter-cluster correlations in the cluster component $\ket{\He(P^{3B};0_k^+)}$
projected from  $\He(0_k^+)$ wave functions,
\begin{equation}
  \begin{split}
    \ket{\He(P^{3B};0_k^+)} &\equiv \frac{P^{3B}\ket{\He(0_k^+)}}{\braket{\He(0_k^+)|P^{3B}|\He(0_k^+)}^{1/2}},   \\
    P^{3B} &= \sum_{i} \ket{\tilde{\Psi}_i^{3B}}\bra{\tilde{\Psi}_i^{3B}},
  \end{split}
\end{equation}
where  $P^{3B}$ is the projection operator given by an orthonormal basis set $\{\tilde \Psi^{3B}_i\}$ constructed from basis wave functions $\{ \ket{P^{0+} \alpha+{}^2n+{}^2n; \rho_i, \gamma_i, \theta_i} \}$.
Note that the cluster components account for more than $90$~\% of $\He(0^+_{1,2})$.
We measure the inter-cluster distances in the cluster component $\ket{\He(P^{3B};0_k^+)}$ by observing pairs of two nucleons with specific spins (say spin-up), and define the distance distributions for $d$ between $\alpha$ and ${}^2n$ clusters and between two ${}^2n$ clusters as
\begin{equation}    \label{dist.dens.8he}
  \begin{split}
    \rho_{\aton}(d=r_{NN}) &\equiv \rho_{p\uparrow n\uparrow}(r_{NN}) \\
    \rho_{\nton}(d=r_{NN}) &\equiv \rho_{n\uparrow n\uparrow}(r_{NN}) - \rho_{p\uparrow n\uparrow}(r_{NN}),
  \end{split}
\end{equation}
\begin{widetext}
  \begin{equation}
   \begin{split}
    \rho_{p\uparrow n\uparrow}(r_{NN}) = \sum_{\substack{i\in \text{proton} \\ j\in \text{neutron}}} \Braket{\He(P^{3B};0_k^+)|\delta(\left| \bm{r}_i-\bm{r}_j \right|-r_{NN}) \frac{P({}^3O)}{3}|\He(P^{3B};0_k^+)} , \\
    \rho_{n\uparrow n\uparrow}(r_{NN}) = \sum_{i,j\in \text{neutron}} \Braket{\He(P^{3B};0_k^+)|\delta(\left|\bm{r}_i-\bm{r}_j \right|-r_{NN}) \frac{P({}^3O)}{3} |\He(P^{3B};0_k^+)},
  \end{split}
  \end{equation}
\end{widetext}
where the operator $\frac{P({}^3O)}{3}$ is inserted to extract the odd-parity component of up-spin 2-nucleon
$(N\uparrow N\uparrow)$ pairs, and $\rho_{p\uparrow n\uparrow}(r_{NN})$ is
equivalent to that of neutron-neutron pairs between the $\alpha$ and ${}^2n$ clusters.
Then, the $\nton$ distance distribution is calculated by subtracting $\rho_{p\uparrow n\uparrow}(r_{NN})$ from $\rho_{n\uparrow n\uparrow}(r_{NN})$.
Note that the integrals of the 2-nucleon densities are equal to the corresponding numbers of pairs, $\int r^2 \rho_{p\uparrow n\uparrow}(r)\,dr=2$ and $\int r^2 \rho_{n\uparrow n\uparrow}(r)\,dr=3$, contained in $\ket{\He(P^{3B};0_k^+)}$.
Furthermore, the integrals of the inter-cluster distance distributions defined in Eq.~\eqref{dist.dens.8he} are equal to the corresponding pair numbers as
$N_{\aton} = \int dr \,r^2\rho_{\aton}(r)=2$ and $N_{\nton}=\int dr \,r^2 \rho_{\nton}(r)=1$.

\subsubsection{$\aton$ and $\nton$ correlations in $\alpha+{}^2n+{}^2n$ structures of $\He$}
We calculate the mean distances $\braket{d}$, their fluctuations $\delta d$, and r.m.s. distances $\sqrt{\braket{d^2}}$ as
\begin{equation}  \label{NNdist}
  \begin{split}
  &\braket{d}= \frac{\int r^2 dr \, r \rho_{C_1\text{-} C_2}(r)}{\int r^2 dr \, \rho_{C_1\text{-} C_2}(r)}, \\
  &\delta d = \left(  \frac{\int r^2 dr \, (r-\braket{d})^2 \rho_{C_1\text{-} C_2}(r)}{\int r^2 dr \, \rho_{C_1\text{-} C_2}(r)} \right)^{1/2}, \\
  &\sqrt{\braket{d^2}}= \left( \frac{\int r^2 dr \, r^2 \rho_{C_1\text{-} C_2}(r)}{\int r^2 dr \, \rho_{C_1\text{-} C_2}(r)} \right)^{1/2}.
  \end{split}
\end{equation}
The calculated mean values of $\He(0_{1,2}^+)$ obtained by the original and balanced calculations are listed in Table~\ref{tab:rhoHe}.

\begin{table}
  \centering
  \caption{Expectation values of the r.m.s distance, mean distance and fluctuation of the mean distance for the $\aton$ and $\nton$ inter-clusters in the original and balanced $\He(0_{1,2}^+)$.}
  \label{tab:rhoHe}
  \begin{ruledtabular}
    \begin{tabular}{ccccc}
      state & channel & $\sqrt{\braket{d^2}}$~(fm) & $\braket{d}$~(fm) & $\delta d$ (fm)\\
      \hline
      orig. $0_1^+$ & $\aton$ & 4.44 & 4.16 & 1.56\\
      orig. $0_1^+$ & $\nton$ & 5.28 & 4.95 & 1.82\\
      orig. $0_2^+$ & $\aton$ & 6.42 & 5.84 & 2.66\\
      orig. $0_2^+$ & $\nton$ & 8.49 & 8.01 & 2.81\\
      bal. $0_1^+$ & $\aton$ & 4.10 & 3.87 & 1.36\\
      bal. $0_1^+$ & $\nton$ & 4.08 & 3.80 & 1.48\\
      bal. $0_2^+$ & $\aton$ & 6.54 & 6.10 & 2.36\\
      bal. $0_2^+$ & $\nton$ & 7.30 & 6.73 & 2.83\\
    \end{tabular}
  \end{ruledtabular}
\end{table}

\begin{figure*}
  \centering
  \includegraphics[width=15cm]{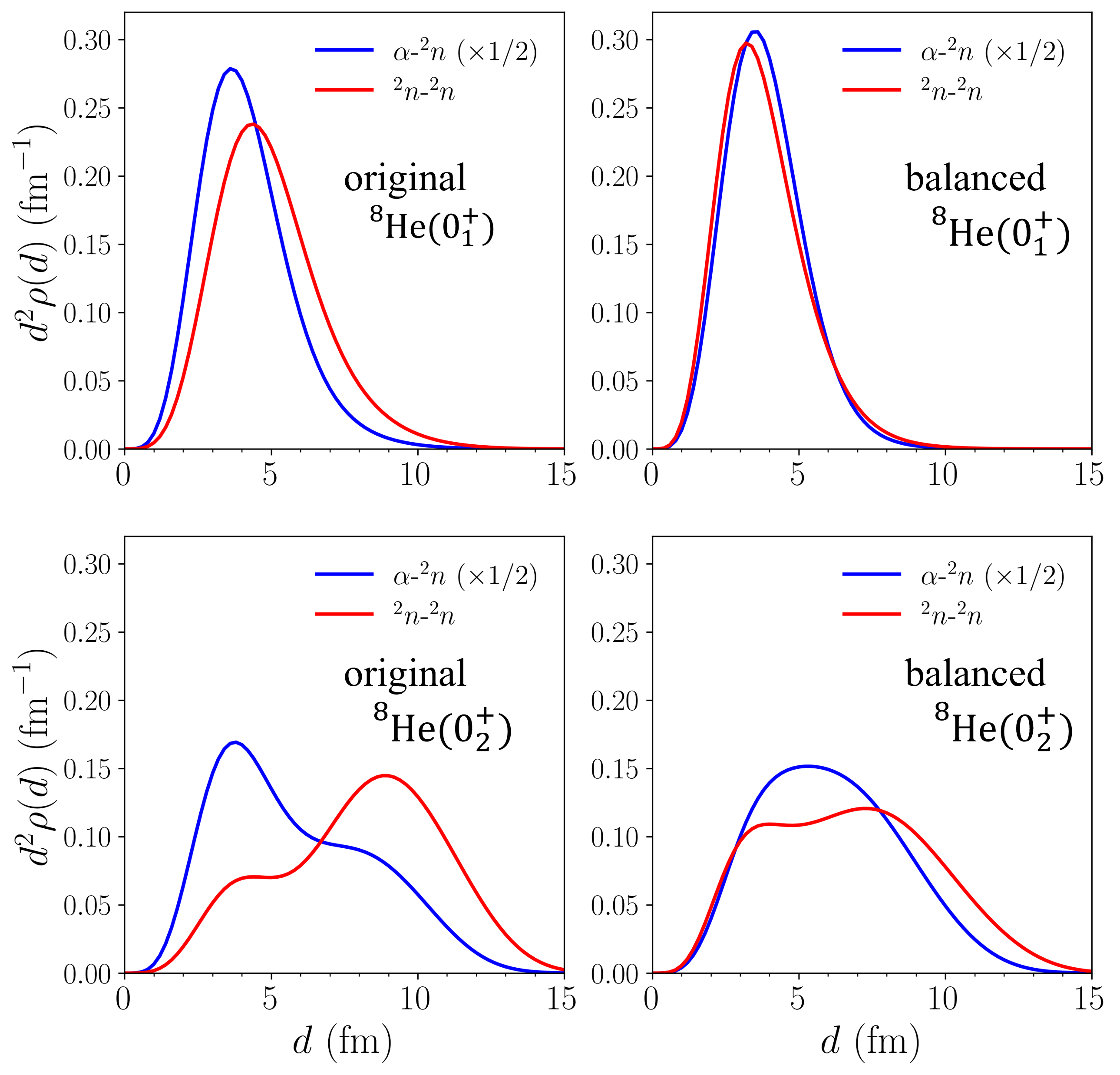}
  \caption{Distance distributions of $\aton$ and $\nton$ clusters in original and balanced $\He(0_{1,2}^+)$.
  For $\aton$ distributions, averaged versions $d^2\rho_{\aton}/2$ are shown.
  The results of the $0_1^+$ and $0_2^+$ states of original $\He$ are placed on the left column of upper and lower panels, and those of balanced $\He$ are arranged on the corresponding right panels.
  $\aton$ and $\nton$ distributions are shown by blue and red lines, respectively.
  }
  \label{fig:d2rho1}
\end{figure*}

Distributions of inter-cluster distances of $\aton$ and $\nton$ in $\He(0_1^+)$ and $\He(0_2^+)$ are plotted as functions of $d$ in Fig.~\ref{fig:d2rho1},
where the averaged values $\rho_{\aton}/2$ of two $\aton$ pairs are compared with $\rho_{\nton}$.

In the original $\He(0_1^+)$, the mean values of the $\aton$ and $\nton$ distances are $4.16$~fm and $4.95$~fm with relatively small fluctuation,
and $3.87$~fm and $3.80$~fm in the balanced $\He(0_1^+)$.
As shown in the top-left panel of Fig.~\ref{fig:d2rho1}, distributions $d^2\rho_{C_1,C_2}(d)$ of $\aton$ and $\nton$ distances show narrow peaks at the corresponding mean values.
These results indicate that positions of three clusters in $\He(0_1^+)$ are well localized at a compact triangle configuration.
For the physical ground state of $\He$ obtained by the original calculation, the $\nton$ distribution is pushed outward compared with the $\aton$ distribution because of the repulsive nature of the original $\nton$ interaction.
As a result, the opening angle between two ${}^2n$ clusters is estimated to be $\sim \ang{73}$, slightly larger than
$\sim \ang{60}$ for the equilateral triangle.

On the other hand, in the $\He(0_2^+)$, the mean values of the $\aton$ and $\nton$ distances are remarkably large because of spatially developed cluster structures in $\He(0_2^+)$ in both cases of the original and balanced calculations.
In the original $\He(0_2^+)$, the $\aton$ distance distribution $r^2\rho_{\alpha\text{-}{}^2n}$ (blue line in the left bottom panel of Fig.~\ref{fig:d2rho1}) exhibits a sharp peak at $d\sim 4$~fm and a broad bump with a long tail in the outer region of $d\gtrsim 6$~fm corresponding to the two component mixing of $\alpha+{}^2n+{}^2n$ and $(\alpha+{}^2n)+{}^2n$.
The former, the sharp peak at $d\sim 4$~fm, indicates the strong $\aton$ correlation, i.e., the $\alpha+{}^2n$ cluster formation, and is mainly contributed by the 2-body $(\alpha+{}^2n)+{}^2n$ cluster component.
The latter, the broad bump in the outer region $d\gtrsim 6$~fm, contains both contributions from the 2-body $(\alpha+{}^2n)+{}^2n$ cluster component and 3-body $\alpha+{}^2n+{}^2n$ cluster-gas components.

The features of the inter-cluster correlations in $\He(0_2^+)$ drastically change from the original $\He$ calculation to the balanced one after the interaction changing.
In the balanced $\He(0_2^+)$ state, the $\aton$ distance distribution $r^2\rho_{\alpha\text{-}{}^2n}$ (blue line in the left bottom panel of Fig.~\ref{fig:d2rho1}) no longer shows a sharp peak and broad bump structure. Instead, the peak at $d\sim 4$~fm for the $(\alpha+{}^2n)+{}^2n$ cluster component disappears and its amplitude shifts toward outer region merging with the outer bump.
As a result, $r^2\rho_{\alpha\text{-}{}^2n}$ exhibits a single broad peak around $d\gtrsim 5$~fm with a long outer tail.
This result indicates that the 2-body $(\alpha+{}^2n)+{}^2n$ cluster component is suppressed, that is, the $(\alpha+{}^2n)$ formation is suppressed in the calculation with balanced inter-cluster interactions.
Namely, the balanced $\He(0_2^+)$ approaches the cluster-gas limit of $\alpha + {}^2n + {}^2n$.

\subsubsection{decomposition of cluster components}
As discussed above, the original $\He(0^+_2)$ results of the distributions and mean values of the $\aton$ and ${\nton}$ distances clearly exhibit the spatial $\aton$ correlation;
the narrow peak of $\rho_{\aton}$ in the original $\He(0^+_2)$ is a signal of the compact
($\alpha$+${}^2n$)-cluster formation and indicates the mixing of the 2-body cluster component. In contrast, in the balanced $\He(0^+_2)$ results, such the signal of the ($\alpha$+${}^2n$)-cluster formation vanishes.
We perform a further detailed analysis of the
inter-cluster distributions to decompose into kinds of cluster components.

The 3-body cluster-gas state of the $0^+_2$ state is characterized by a dilute cluster-gas in the outer part but still contains ground state component in the inner part because of orthogonality to  the ground state as in $\C(0^+_2)$,
and therefore, the cluster-gas limit of $\He(0^+_2)$ is also expected to contain two components; the outer part of the dilute cluster-gas, where two neutrons distribute in the outer region around the alpha cluster and are freely moving like a gas,
and inner part of the compact triangle configuration due to the orthogonality to $\He(0^+_1)$.
On the other hand, the 2-body ($\alpha$+$^2n$)+$^2n$ cluster component of $\He(0^+_2)$ has the compact ($\alpha$+$^2n$) cluster with a surrounding neutron in the outer region.

In the following discussions, we separate the region of $\aton$ distance into two parts of the inner and outer regions at a border $d_{}=d_{\textrm{b}}$, for simplicity.
The inner region is the compact area as small as the size of the ground state, corresponding to dineutrons nearby the $\alpha$ cluster, whereas the outer region is the case of dineutrons far from the $\alpha$ cluster.
We also separate the distributions of the $\alpha$-$^2n$ distance into the inner and outer parts,
$\rho^{\textrm{(in)}}_{\aton}$ and $\rho^{\textrm{(out)}}_{\aton}$, at the border $d_{}=d_{\textrm{b}}$. Here we choose $d_{\textrm{b}}=5$ fm to cover the inner area $d_{}<d_{\textrm{b}}$ as small as the $d^2\rho_{\aton}$ distribution of the ground state.

Let us consider contributions from three components, the 2-body cluster, compact triangle, and dilute cluster-gas ones, to the two parts
$\rho^{\textrm{(in)}}_{\aton}$ and
$\rho^{\textrm{(out)}}_{\aton}$.

\begin{figure*}
  \centering
  \includegraphics[width=15cm]{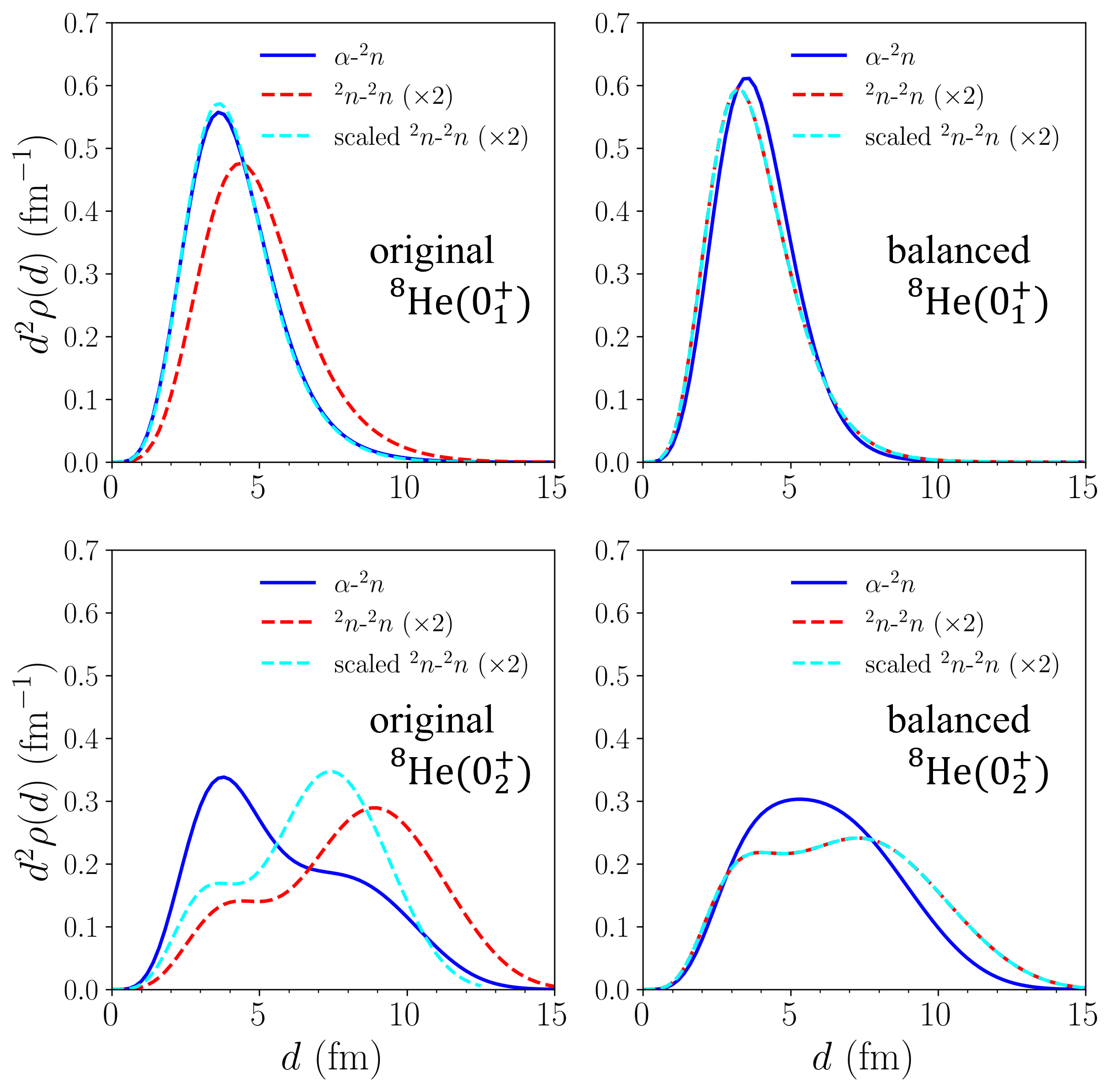}
  \caption{Distance distributions as Fig.~\ref{fig:d2rho1}, but unaveraged $\aton$ distributions are shown instead of the averaged ones and $\nton$ scaled distributions are added.
  The results of unaveraged $\aton$ distributions and $\nton$ scaled distributions $2s^3{\rho}_{\nton}(sd)$ are plotted by dotted red and cyan lines, respectively.
  For the detailed definitions, refer to the main text.
  }
  \label{fig:d2rho}
\end{figure*}

 Considering that the 2-body contains a nearside $^2n$ and a far-side $^2n$
due to the compact ($\alpha{\textrm{+}}^2n$)-cluster formation,
the compact triangle has two nearside $^2n$s, and
the dilute cluster-gas has two far-side $^2n$s, we get decompositions
 of each part as,
\begin{equation}
  \begin{split}
   \rho_{\aton}^{\textrm{(in)}}=\rho_{\aton}^{\textrm{(in)}}[\textrm{2B}]+\rho_{\aton}[\textrm{tri}], \\
   \rho_{\aton}^{\textrm{(out)}}=\rho_{\aton}^{\textrm{(out)}}[\textrm{2B}]+\rho_{\aton}[\textrm{gas}], \\
  \end{split}
\end{equation}
where
$\rho^{\textrm{(in/out)}}_{\aton}[\textrm{2B}]$,
$\rho_{\aton}[\textrm{tri}]$, and
$\rho_{\aton}[\textrm{gas}]$ indicate contributions from the 2-body, triangle, and dilute gas components.
We assume that $\aton$ distributions can be approximated by scaling ${\nton}$ distributions as
\begin{equation}  \label{rhotri_ap}
   \rho_{\aton}[\textrm{tri}](d_{}) \approx 2 {\rho}^{\textrm{(in)}}_{s(\nton)}(s;d_{})
\end{equation}
with
\begin{equation}
  {\rho}^{\textrm{(in)}}_{s(\nton)}(s;d_{}) \equiv s^3\rho_{\nton}(sd_{}), \qquad sd_{}<sd_{\textrm{b}},
\end{equation}
where ${\rho}_{s(\nton)}$ is scaled distribution functions of ${\nton}$ distributions
and $s$ is the scaling factor for the distance scaling
from the $\aton$ to ${\nton}$ distances,
$d_{}\to sd_{}$.
As a result, we get the following approximation,
 \begin{equation} \label{rhoin2B}
  \rho^{\textrm{(in)}}_{\aton}[\textrm{2B}](d_{})\approx
 \rho^{\textrm{(in)}}_{\alpha\textrm{-}^2n}(d_{})-2{\rho}^{\textrm{(in)}}_{s(\nton)}(s;d_{}).
 \end{equation}
By comparing $\rho^{\textrm{(in)}}_{\aton}$ with $2{\rho}^{\textrm{(in)}}_{s(\nton)}$, we evaluate the 2-body component through the nearside dineutron contribution to the inner $\aton$ distribution. The original and balanced results of $2\rho_{\aton}$ and $2{\rho}_{s(\nton)}$ are shown in Fig.~\ref{fig:d2rho} together with unscaled ${\nton}$ distributions $\rho_{\nton}$.

Here, we choose the scaling factor $s$ for $d_{} \to sd_{}$ of ${\rho}_{s(\nton)}$ to be  the ratio of the mean value of the $\aton$ distance to that of the ${\nton}$ distance in the ground state $\He(0^+_1)$, and adopt $s=1.2$ and $s=1.0$ for the original and balanced calculations from the values 4.95/4.16 and 3.80/3.87 in Table~\ref{tab:rhoHe}, respectively.

In the original result of $\He(0^+_2)$ (bottom-left panel of Fig.~\ref{fig:d2rho}),
 $\rho_{\aton}$ significantly exceeds $2\rho_{s(\nton)}(s;d_{})$ in the inner region.
This excess is nothing but the distribution
$\rho^{\textrm{(in)}}_{\aton}[\textrm{2B}]$
of the nearside dineutron in the 2-body component as described in Eq.~\eqref{rhoin2B},
and its narrow peak  at $d\approx 4$~fm indicates the formation of
the $\alpha$+$^2n$ cluster as compact as the size of the $\He$ ground state.
By integrating this excess in the inner region ($d<d_\textrm{b}$),
we measure the number
of the nearside neutron of the 2-body component,
\begin{equation}
  \begin{split}
    N^{\textrm{(in)}}_{\aton}[\textrm{2B}]
  \equiv\int_0^{d_{\textrm{b}}} dr r^2 \rho_{\aton}[\textrm{2B}](r) \\
  \approx \int_0^{d_{\textrm{b}}} dr r^2
  \{ \rho_{\aton}(r) -2{\rho}_{s(\nton)}(s;r) \},
  \end{split}
\end{equation}
which corresponds to the fraction of the 2-body component
contained in the $\He(0^+_2)$ state.
Using the calculated results of $\rho_{\aton}$ and ${\rho}_{\nton}$, we obtain $N^{\textrm{(in)}}_{\aton}[\textrm{2B}]\approx 0.4$, i.e.,
 40\% mixing of the 2-body cluster component in the original $\He(0^+_2)$.

On the other hand, in the balanced $\He(0^+_2)$,
the excess of $\rho_{\aton}$ from $2{\rho}_{s(\nton)}(s;d_{})$ in the inner region is reduced, and
the narrow peak at $d=4$~fm disappears, indicating that the $\alpha$+$^2n$ cluster is no longer formed.
We calculate
the integration of the excess in the inner region, and obtain the value
$N^{\textrm{(in)}}_{\aton}[\textrm{2B}]\approx 0.1$ corresponding to 10\% of the 2-body cluster mixing in the balanced $\He(0^+_2)$, which indicates
that the 2-body cluster mixing is strongly suppressed.

From these analyses of inter-cluster distance distributions,
the significant mixing of the 2-body cluster component in the original $\He(0^+_2)$ and its suppression in the balanced $\He(0^+_2)$ are clearly detected.

\subsection{inter-cluster distance in $\Be$}
We perform similar analysis of inter-cluster correlations in $\Be$. We measure the inter-cluster distances of $\atoa$ and $\aton$ in the $2\alpha+{}^2n$ components $\Be(P^{3B};0_k^+)$ projected from the $\Be(0_k^+)$ wave functions by observing $NN$ pairs in the ${}^3O$ channel, and calculate the inter-cluster distance distributions given as
\begin{equation}
  \begin{split}
    \rho_{\alpha \text{-} \alpha}(d=r_{NN}) &\equiv \rho_{p\uparrow p\uparrow}(r_{NN}) \\
    \rho_{\alpha-{}^2n}'(d=r_{NN}) &\equiv  \rho_{p\uparrow n\uparrow}(r_{NN})
- \rho_{p\uparrow p\uparrow}(r_{NN}),
  \end{split}
\end{equation}
where $\rho_{p\uparrow p\uparrow}(r_{NN})$ is the distribution of the $p\uparrow p\uparrow$ pairs with the distance $r_{NN}$ as
\begin{equation}
  \rho_{p\uparrow p\uparrow}(r_{NN}) = \sum_{i,j\in \text{proton}}\delta(|\bm{r}_i-\bm{r}_j|-r_{NN}) \frac{P({}^3O)}{3}.
\end{equation}
We calculate the mean distances, their fluctuations, and r.m.s. distances of $\atoa$ and $\aton$ for the $\Be(0_{1,2}^+)$ wave functions obtained by the original and balanced calculations, and show the results in Table~\ref{tab:rhoBe}.

\begin{table}
  \centering
  \caption{Same as in Table~\ref{tab:rhoHe}, but for the $\aton$ and $\atoa$ distances of the original and balanced $\Be(0_{1,2}^+)$ calculations.}
  \label{tab:rhoBe}
  \begin{ruledtabular}
    \begin{tabular}{ccccc}
      state & channel & $\sqrt{\braket{d^2}}$~(fm) & $\braket{d}$~(fm) & $\delta d$ (fm)\\
      \hline
      orig. $0_1^+$ & $\atoa$ & 4.49 & 4.25 & 1.44\\
      orig. $0_1^+$ & $\aton$ & 4.44 & 4.17 & 1.51\\
      orig. $0_2^+$ & $\atoa$ & 5.98 & 5.63 & 2.02\\
      orig. $0_2^+$ & $\aton$ & 5.68 & 5.18 & 2.31\\
      bal. $0_1^+$ & $\atoa$ & 4.26 & 4.03 & 1.37\\
      bal. $0_1^+$ & $\aton$ & 4.54 & 4.28 & 1.54\\
      bal. $0_2^+$ & $\atoa$ & 4.97 & 4.65 & 1.75\\
      bal. $0_2^+$ & $\aton$ & 6.79 & 6.28 & 2.59\\
    \end{tabular}
  \end{ruledtabular}
\end{table}

In the original $\Be(0_1^+)$ state, the mean values of the $\atoa$ and $\aton$ distances are $4.25$~fm and $4.17$~fm, respectively, with relatively small fluctuations because of a compact equilateral-triangular cluster structure in the physical ground state of $\Be$.
In the original $\Be(0_2^+)$ state, the mean $\atoa$ distance ($5.63$~fm) is larger than that of $\aton$ ($5.18$~fm) of the $\Be(0_2^+)$ state.
Considering that its fluctuation is not as large as those obtained for $\He(0_2^+)$, the $\atoa$ distance distribution in the $\Be(0_2^+)$ state is localized around the mean distance, indicating that the $\Be(0_2^+)$ state has a stretched $2\alpha$ structure with a dumbbell-like configuration at the distance $\sim 5.63$~fm.
As shown in the right bottom panel of Fig.~\ref{fig:ougi.10Be2} for the dineutron distribution around the $\atoa$ fixed at $D_{\alpha\text{-}\alpha}=6$ fm,
the dineutron in the $\Be(0_2^+)$ distributes in the elongated region around the dumbbell $2\alpha$ structure.
In the results of the balanced $\Be$ calculations, the inter-cluster distances in $\Be(0_1^+)$ indicates again a compact equilateral-triangle configuration.
As for the balanced $\Be(0_2^+)$ state, we found a similar trend of the $\atoa$ distance to the original $\Be(0_2^+)$ result;
a larger mean value of $\Be(0_2^+)$ than the ground state ($4.65$ fm in $\Be(0_2^+)$ and $4.03$ fm in $\Be(0_1^+)$) and a small fluctuation, which indicate again a stretched $2\alpha$ structure with a dumbbell-like configuration at the distance $\sim 4.65$~fm. It is consistent with
the feature of the angular excitation mode in the balanced $\Be(0_2^+)$ state previously discussed
in Sec.~\ref{sec:clus10Be}.
We do not look into details of the $\aton$ distance because the distance distribution provides only an inclusive value averaged over angles and is not suitable for describing
the angular dependence of the dineutron distribution around $2\alpha$.

\subsection{angular dependence of 3-body cluster configurations}  \label{sec:ang10Be}
\begin{figure*}
  \centering
  \includegraphics[width=15cm]{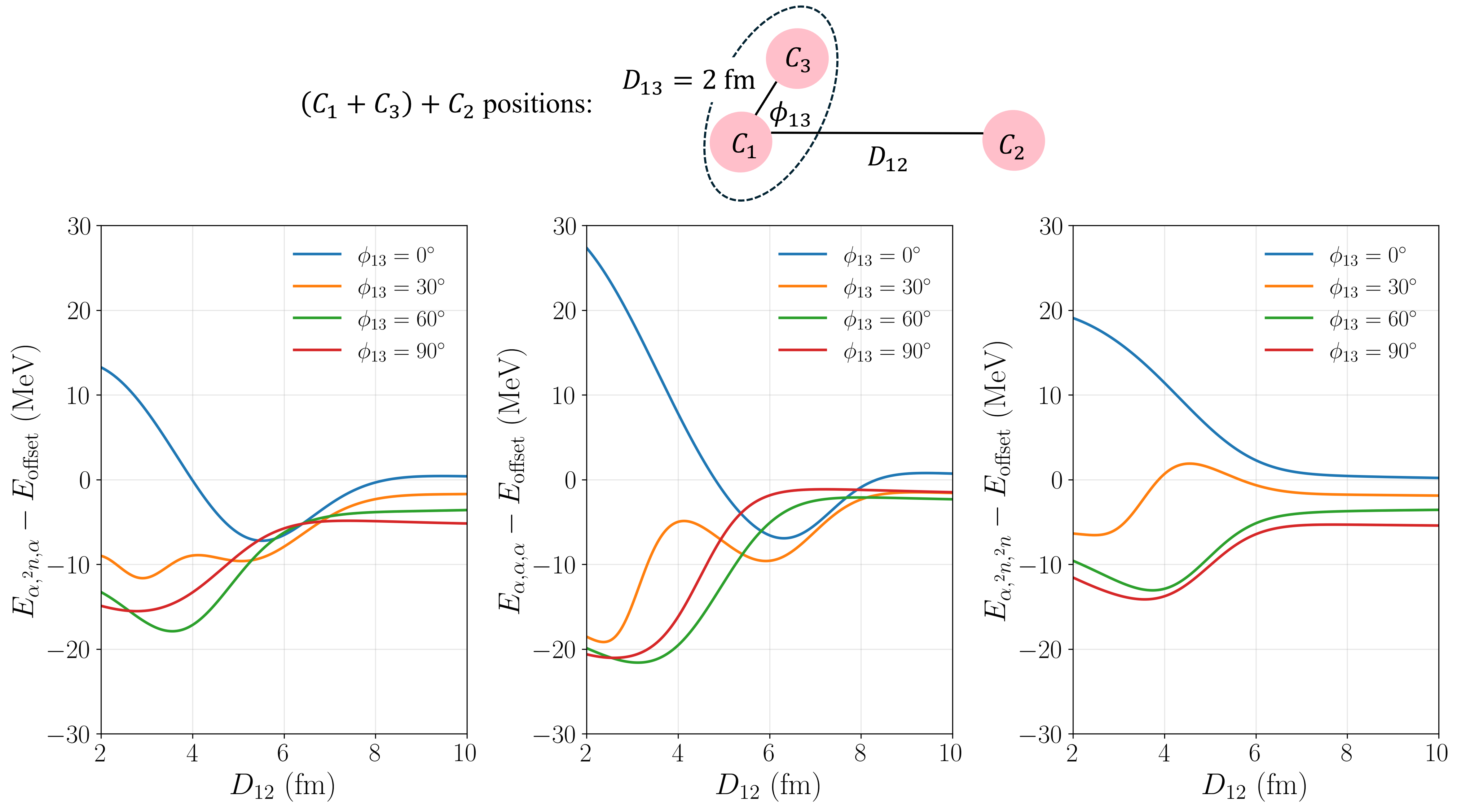}
  \caption{
    Energies of 3-cluster systems $(C_1+C_3)+C_2$ for $(C_1,C_2,C_3)=(\alpha,{}^2n,\alpha), (\alpha,\alpha,\alpha),({}^2n,\alpha,{}^2n)$ are plotted as functions of $D_{12}$ on the left, middle and right panels, respectively.
$D_{13}$ is fixed at $D_{13}=\SI{2}{fm}$ and the results of  $\phi_{13}=0^{\circ},30^{\circ},\cdots,90^{\circ}$ cases are shown in each panels. (See the schematic figure at the top for the cluster positions.)
    The energies are measured from the offset energies $E_{\textrm{offset}}$, which are chosen to be the energies at $D_{12}=15$~fm and $\phi_{13}=0^\circ$ for each system.
  }
  \label{fig:3Bdist}
\end{figure*}

The analyses of $\rho$-fixed calculations and those of dineutron distributions
in Sec.~\ref{sec:res}
suggest a significant difference of the cluster motion in the angular direction of
the $\Be(0^+_2)$ from those
of $\C(0^+_2)$ and  $\He(0^+_2)$.
The 3$\alpha$ and $\alpha+{}^2n+{}^2n$ systems
favor the radial excitation, whereas the $2\alpha+{}^2n$ system shows
an angular excitation mode rather than the radial excitation one.
To clarify the reason for this difference,
we here discuss the angle dependence of energies of 3-cluster systems defined as
\begin{widetext}
  \begin{equation} \label{3Bangle}
    E_{C_1,C_2,C_3}(D_{12},D_{13},\phi_{13})= \frac{\braket{C_1+C_2+C_3;D_{12},D_{13},\phi_{13}|HP^{0+}|C_1+C_2+C_3;D_{12},D_{13},\phi_{13}}}{\braket{C_1+C_2+C_3;D_{12},D_{13},\phi_{13}|P^{0+}|C_1+C_2+C_3;D_{12},D_{13},\phi_{13}}} - \sum_{i=1}^3 E_{C_i},
  \end{equation}
\end{widetext}
which can be decomposed into the kinetic part $T_{C_1,C_2,C_3}$, nuclear potential part $V_{C_1,C_2,C_3}$, and Coulomb potential part $V^{coul}_{C_1,C_2,C_3}$ as,
\begin{eqnarray}
&  E_{C_1,C_2,C_3}(D_{12},D_{13},\phi_{13}) =T_{C_1,C_2,C_3}(D_{12},D_{13},\phi_{13})\nonumber \\
  &+ V_{C_1,C_2,C_3}(D_{12},D_{13},\phi_{13}) + V^{coul}_{C_1,C_2,C_3}(D_{12},D_{13},\phi_{13}).\nonumber \\
\end{eqnarray}
We fix the $C_1$-$C_3$ distance at $D_{13}=2$~fm
and calculate energies as functions of the opening angle
$\phi_{13}$ and the $C_1$-$C_2$ distance $D_{12}$.
As shown in schematic figures in Fig.~\ref{fig:3Bdist},
the angle $\phi_{13}$ corresponds to the orientation of
the $(C_1+C_3)$ cluster with respect to the direction of the
$C_2$ cluster located at the distance $D_{12}$.

\begin{figure*}
  \centering
  \includegraphics[width=15cm]{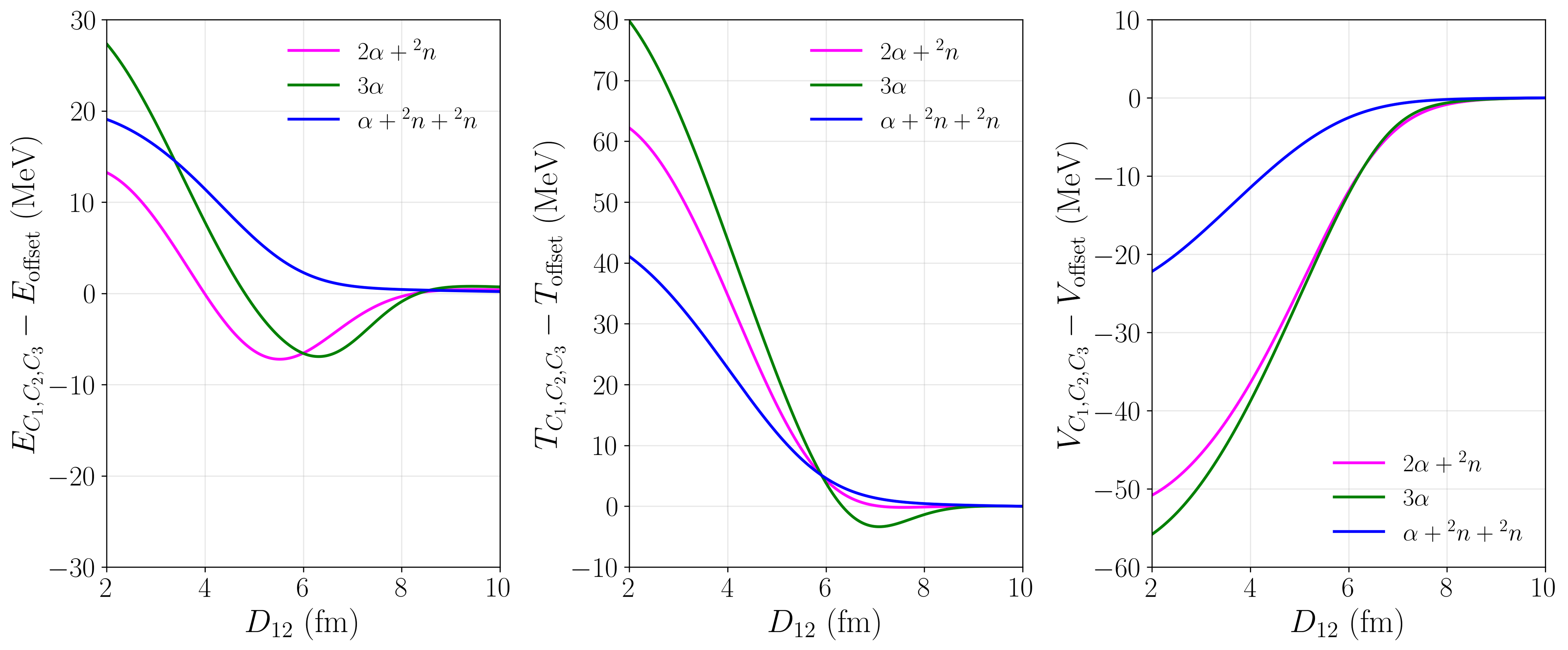}
  \caption{$E_{C_1,C_2,C_3}$, $T_{C_1,C_2,C_3}$, and $V_{C_1,C_2,C_3}$ at
 $\phi_{13}=0^{\circ}$ for the linear configurations, in the left, middle, and right panels, respectively.
In each panel, energies for $(C_1,C_2,C_3)=(\alpha,{}^2n,\alpha), (\alpha,\alpha,\alpha),({}^2n,\alpha,{}^2n)$
 measured from the offset energies $E_{\textrm{offset}}$, $T_{\textrm{offset}}$, and $V_{\textrm{offset}}$
at $D_{12}=15$~fm are compared by magenta, blue and green lines, respectively.
$D_{13}$ is fixed at $D_{13}=\SI{2}{fm}$.
}
  \label{fig:3Bdistphi0}
\end{figure*}

The energies $E_{C_1,C_2,C_3}$, $T_{C_1,C_2,C_3}$, and $V_{C_1,C_2,C_3}$ are calculated for $(C_1,C_2,C_3)=(\alpha,{}^2n,\alpha),(\alpha,\alpha,\alpha)$, and $({}^2n,\alpha,{}^2n)$ with the original interaction.
Figure~\ref{fig:3Bdist} shows the energies $E_{C_1,C_2,C_3}$ at $\phi_{13}=0^{\circ}$, $30^{\circ}$, $60^{\circ}$, and $90^{\circ}$ plotted as functions of $D_{12}$~fm, and Fig.~\ref{fig:3Bdistphi0} exhibits the results of $E_{C_1,C_2,C_3}$, $T_{C_1,C_2,C_3}$, and $V_{C_1,C_2,C_3}$
at $\phi_{13}=0^{\circ}$ for the linear configuration.
In these figures, we set the energy offset at $D_{12}=15$~fm and
$\phi_{13}=0^\circ$, and plot the energies measured from
the offset energies.

As shown in Figs.~\ref{fig:3Bdist}~and~\ref{fig:3Bdistphi0}, the energy curves strongly depend on the angle $\phi_{13}$ in all systems indicating the
significant dependence on the $(C_1+C_3)$-cluster orientation.
The $\phi_{13}=60\sim 90^\circ$ energy curves are
lowest and their energy minima correspond to the compact triangle
configuration of the main components of the $0^+_1$ states.
Compared with the $R$-fixed energies shown in Fig.~\ref{fig:adbt.comp} and Fig.~\ref{fig:adbt.10Be},
the $\phi_{13}=60^\circ$ and $\phi_{13}=0^\circ$ energy curves
show similar trends to the $\nu=1$ and $\nu=2$ energy curves,
respectively. It indicates that the $\nu=2$ mode is the angular
excitation corresponding to the $\phi_{13}=0^\circ$ configuration for
the linear structure. In particular,
the $\phi_{13}=0^\circ$ energy curve for $\Be$ shown
in the left panel of Fig.~\ref{fig:3Bdist} exhibits
the property qualitatively similar to that of the $\nu=2$
curve of the $R$-fixed energy in Fig.~\ref{fig:adbt.10Be}; the $\phi_{13}=0^\circ$
energy has the minimum at $D_{12}\sim 5$~fm, which corresponds to the
energy minimum of the $\nu=2$ energy curve at the hyper radius $R\sim 7$~fm.
This energy minimum at $D_{12}\sim 5$~fm of the $\phi_{13}=0^\circ$ energy curve
in $\Be$ is deep enough to stabilize the linear-chain configuration and
mix this component into the $\Be(0^+_2)$.

On the other hand, it is not the case
in $\C$ and $\He$ systems.
In $\C$, the $\phi_{13}=0^\circ$ energy curve shows an energy pocket
at $D_{12}=6\sim 7$~fm larger distance than  $\Be$, and furthermore,
its depth is much higher compared with the $\phi_{13}=60^\circ$ energy minimum.
As a result of the larger distance and higher energy of the $\phi_{13}=0^\circ$ energy minimum,  the linear-chain configuration is not favored in the $\C(0^+_2)$ state.
In the case of $\He$, there is no energy minimum in the $\phi_{13}=0^\circ$ energy curve, indicating that the linear-chain configuration of $\alpha+{}^2n+{}^2n$ is unstable.

These differences in the $\phi_{13}=0^\circ$ energy behavior
between $\Be$, $\C$, and $\He$ systems are understood
by system dependence of kinetic and nuclear potential energy contributions,
for which microscopic effects in 3-cluster systems play important roles.
The energies at $\phi_{13}=0^\circ$ for the linear configuration of three systems
are compared in Fig.~\ref{fig:3Bdistphi0};
the left, middle, and right panels show the total energies $E_{C_1,C_2,C_3}$, the kinetic energies $T_{C_1,C_2,C_3}$, and the potential energies $V_{C_1,C_2,C_3}$
respectively.
In $\Be$, the energy pocket at  $D_{12}\sim 5$~fm is produced by the
moderate kinetic repulsion at the short distance and
the deep potential energy. In $\C$, the kinetic repulsion is much higher than the
$\Be$ case because of the stronger Pauli repulsion
between $C_2=\alpha$ and $C_3=\alpha$ clusters
in the $3\alpha$ system than that between $C_2={}^2n$ and $C_3=\alpha$ clusters
in the $2\alpha+{}^2n$ system.
On the other hand, the potential energy $V_{C_1,C_2,C_3}$ in $\C$
is approximately the same as that of $\Be$. As a result, the
$\phi_{13}=0^\circ$ energy pocket shifts toward the large distance region of
$D_{12}=6\sim7$~fm, and therefore, the linear-chain structure of
$\C$ is not as robust as the tightly bonded linear-chain structure in $\Be$.

It should be pointed out that the  microscopic effects
specific to 3-cluster systems beyond 2-cluster system give essential
contribution to the potential energy
$V_{C_1,C_2,C_3}$ through the internal energy loss of clusters from the Pauli blocking effect.
Provided that the potential energies of $2\alpha+{}^2n$ and $3\alpha$ systems are
given by a sum of the inter-cluster potential energies of three pairs of
clusters $C_1$-$C_2$, $C_1$-$C_3$, and $C_2$-$C_3$, it is expected that
$V_{C_1,C_2,C_3}$ of $2\alpha+{}^2n$ is higher than that of  $3\alpha$.
However, the results of $V_{C_1,C_2,C_3}$
in 3-cluster systems contradict this naive expectation;
$V_{C_1,C_2,C_3}$ of $2\alpha+{}^2n$ is almost the same as that of $3\alpha$,
meaning that the $2\alpha+{}^2n$ system gains additional potential
energy. This is a microscopic effect
in 3-cluster systems beyond 2-cluster systems
due to the system-dependence of the Pauli blocking effect.

In $\He$, the kinetic repulsion is weaker than $\Be$ and $\C$ systems, but
the  potential energy attraction is small and not enough to produce
an energy minimum. As a result, the total energy of the linear
$\alpha+{}^2n+{}^2n$ system in $D_{12}<10$~fm is always higher
than the asymptotic energy at $D_{12}=15$~fm
as expected from the repulsive nature of the $^2n$-$^2n$ inter-cluster energy.

\section{summary}
\label{sec:sum}
We investigated the  cluster structures of $\alpha+{}^2n+{}^2n$ in $\He$ and
$2\alpha+{}^2n$ in $\Be$, and compared them with those of $3\alpha$ in  $\C$
focusing on  the 3-cluster gas structure in excited $0^+$ states.
Detailed analyses were performed to understand the emergence mechanism of
the 3-body cluster-gas states in the excited $0^+$ states.

To understand the features of  $\alpha+{}^2n+{}^2n$ clustering in $\He$,
we discussed the subsystem energies, i.e.,  inter-cluster $\nton$ and $\aton$  energies
and showed that $\aton$ energy is more attractive than the $\atoa$ energy, whereas $\nton$ energy is repulsive.
This unbalance between the $\nton$ and $\aton$ interactions is the origin of
the 2-body-like $(\alpha+{}^2n)+{}^2n$ mixing in the 3-body cluster-gas state of  $\He(0^+_2)$.
Analysis of  the inter-cluster energies revealed important roles of
microscopic effects of nucleon degrees of freedom, that is, Pauli blocking effects between clusters through
the kinetic energy loss and  internal potential energy loss of clusters.

We also performed test calculations by artificially
changing  the nuclear force parameters to control the $\aton$ and $\nton$ interactions and
examine how the cluster structures vary in the balanced case of the $\aton$ and $\nton$ interactions.
It was found that the 2-body
 $(\alpha+{}^2n)+{}^2n$ mixing is suppressed and the $\He(0^+_2)$ state approaches
the ideal 3-body cluster-gas state in the balanced case.
An insight from the systematic studies of three systems, $\He$, $\Be$, and $\C$
is that the balance between inter-cluster interactions is an essential condition for the emergence of
the 3-body cluster-gas states in the $0^+_2$ states, but an exception is the $\Be$ case, in which
the bending excitation mode exists in the energy region lower than the radial excitation mode
and the 3-body cluster-gas state does not appear in $\Be(0^+_2)$.

For further detailed analyses of cluster features of $\He$ and $\Be$,
we calculated the inter-cluster distance distributions and showed that the ground states of $\He$ and $\Be$ can be expressed by an isosceles triangle and an equilateral triangle, respectively.
From the inter-cluster distance distributions,  the 2-body  $(\alpha+{}^2n)+{}^2n$ mixing is estimated to be
$\SI{40}{\%}$ in the default $\He(0_2^+)$ state, whereas the balanced $\He(0_2^+)$ contains
only $\SI{10}{\%}$ of the mixing, showing the dominance of the dilute cluster-gas component.

It should be stressed again that the microscopic effects of nucleon degrees of freedom
give important contributions to cluster features of the ground and excited states of
$\He$, $\C$, and $\Be$.
In particular, the Pauli blocking effects play essential roles through the kinetic energy loss and
internal nuclear potential energy loss of clusters. These microscopic effects provide
non-trivial contributions to 3-cluster systems beyond non-microscopic 3-body systems consisting of
structureless particles.

To understand the universality of cluster-gas states, it will be necessary to investigate
multi-cluster systems of $\alpha$ and ${}^2n$ clusters in heavier nuclei. It is also a
challenging issue to investigate dineutron clustering in other neutron-rich systems such as
${}^{12}\mathrm{Be}$ and neutron-rich C and O isotopes.

\begin{acknowledgments}
This work was supported by Grants-in-Aid of the Japan Society for the Promotion of Science (Grant No. 22K03633) and JST SPRING (Grant No. JPMJSP2110).
The numerical calculations were carried out on Yukawa-21 at YITP in Kyoto University.
\end{acknowledgments}
\bibliographystyle{apsrev4-2}
\bibliography{mainbib}
\end{document}